\documentclass[twocolumn]{aastex62}

\usepackage{color}
\usepackage{amsmath} 
\usepackage{graphicx}

\newcommand{\nhits}{54 } 
\newcommand{\nhours}{4348 }

\newcommand{\data}{\tilde d}
\newcommand{\model}{\tilde h}
\newcommand{\residual}{\tilde r}
\newcommand{\params}{{\vec \lambda}}
\newcommand{\n}{\tilde n}
\newcommand{\Cij}{C_{ij}}

\newcommand{\invCij}{C_{ij}^{-1}}

\received{\today}
\revised{\today}
\accepted{\today}
\submitjournal{ApJ}

\shorttitle{Micrometeoroid Impacts in \emph{LISA Pathfinder}}
\shortauthors{Thorpe, et al.}

\begin{document}

\title{Micrometeoroid Events in \emph{LISA Pathfinder}}

\correspondingauthor{J.I. Thorpe}
\email{james.i.thorpe@nasa.gov}


\author[0000-0001-9276-4312]{J\,I~Thorpe}
\affiliation{Gravitational Astrophysics Lab, NASA Goddard Space Flight Center, 8800 Greenbelt Road, Greenbelt, MD 20771 USA}

\author{J~Slutsky}
\affiliation{Gravitational Astrophysics Lab, NASA Goddard Space Flight Center, 8800 Greenbelt Road, Greenbelt, MD 20771 USA}

\author{John G. Baker}
\affiliation{Gravitational Astrophysics Lab, NASA Goddard Space Flight Center, 8800 Greenbelt Road, Greenbelt, MD 20771 USA}

\author{Tyson B. Littenberg}
\affiliation{NASA Marshall Space Flight Center,  Huntsville, AL 35812, USA}

\author{Sophie Hourihane}
\affiliation{NASA Marshall Space Flight Center, Huntsville, AL 35812, USA}
\affiliation{Gravitational Astrophysics Lab, NASA Goddard Space Flight Center, 8800 Greenbelt Road, Greenbelt, MD 20771 USA}
\affiliation{University of Michigan}

\author{Nicole Pagane}
\affiliation{Gravitational Astrophysics Lab, NASA Goddard Space Flight Center, 8800 Greenbelt Road, Greenbelt, MD 20771 USA}
\affiliation{Johns Hopkins University}

\author{Petr Pokorny}
\affiliation{Code 674, NASA Goddard Space Flight Center, 8800 Greenbelt Road, Greenbelt, MD 20771 USA}
\affiliation{Catholic University of America}

\author{Diego Janches}
\affiliation{Code 674, NASA Goddard Space Flight Center, 8800 Greenbelt Road, Greenbelt, MD 20771 USA}
\nocollaboration

\collaboration{(The LISA Pathfinder Collaboration)}
\author{M~Armano}\affiliation{European Space Astronomy Centre, European Space Agency, Villanueva de la Ca\~{n}ada, 28692 Madrid, Spain}
\author{H~Audley}\affiliation{Albert-Einstein-Institut, Max-Planck-Institut f\"ur Gravitationsphysik und Leibniz Universit\"at Hannover, Callinstra{\ss}e 38, 30167 Hannover, Germany}
\author{G~Auger}\affiliation{APC, Univ Paris Diderot, CNRS/IN2P3, CEA/lrfu, Obs de Paris, Sorbonne Paris Cit\'e, France}
\author{J~Baird}\affiliation{High Energy Physics Group, Physics Department, Imperial College London, Blackett Laboratory, Prince Consort Road, London, SW7 2BW, UK}
\author{M~Bassan}\affiliation{Dipartimento di Fisica, Universit\`a di Roma ``Tor Vergata'',  and INFN, sezione Roma Tor Vergata, I-00133 Roma, Italy}
\author{P~Binetruy}\altaffiliation{Deceased}\affiliation{APC, Univ Paris Diderot, CNRS/IN2P3, CEA/lrfu, Obs de Paris, Sorbonne Paris Cit\'e, France}
\author{M~Born}\affiliation{Albert-Einstein-Institut, Max-Planck-Institut f\"ur Gravitationsphysik und Leibniz Universit\"at Hannover, Callinstra{\ss}e 38, 30167 Hannover, Germany}
\author{D~Bortoluzzi}\affiliation{Department of Industrial Engineering, University of Trento, via Sommarive 9, 38123 Trento, and Trento Institute for Fundamental Physics and Application / INFN}
\author{N~Brandt}\affiliation{Airbus Defence and Space, Claude-Dornier-Strasse, 88090 Immenstaad, Germany}
\author{M~Caleno}\affiliation{European Space Technology Centre, European Space Agency, Keplerlaan 1, 2200 AG Noordwijk, The Netherlands}
% ...

\vspace*{0.5in}
\collaboration{(The ST7-DRS Operations Team)}
 \author{P~Barela}\affiliation{NASA Jet Propulsion Laboratory, California Institute of Technology, Pasadena, CA 91109 USA}
 \author{C~Cutler}\affiliation{NASA Jet Propulsion Laboratory, California Institute of Technology, Pasadena, CA 91109 USA}
 \author{N~Demmons}\affiliation{Busek Co., Natick, MA 01760 USA}
 \author{C~Dunn}\affiliation{NASA Jet Propulsion Laboratory, California Institute of Technology, Pasadena, CA 91109 USA}
 \author{M~Girard}\affiliation{NASA Jet Propulsion Laboratory, California Institute of Technology, Pasadena, CA 91109 USA}
 \author{O~Hsu}\affiliation{Attitude Control Systems Branch, NASA Goddard Space Flight Center, 8800 Greenbelt Road, Greenbelt, MD 20771 USA}
 \author{S~Javidnia}\affiliation{NASA Jet Propulsion Laboratory, California Institute of Technology, Pasadena, CA 91109 USA}
 \author{I~Li}\affiliation{NASA Jet Propulsion Laboratory, California Institute of Technology, Pasadena, CA 91109 USA}
 \author{P~Maghami}\affiliation{Attitude Control Systems Branch, NASA Goddard Space Flight Center, 8800 Greenbelt Road, Greenbelt, MD 20771 USA}
 \author{C~Marrese-Reading}\affiliation{NASA Jet Propulsion Laboratory, California Institute of Technology, Pasadena, CA 91109 USA}
 \author{J~Mehta}\affiliation{NASA Jet Propulsion Laboratory, California Institute of Technology, Pasadena, CA 91109 USA}
 \author{J~O'Donnell}\affiliation{Attitude Control Systems Branch, NASA Goddard Space Flight Center, 8800 Greenbelt Road, Greenbelt, MD 20771 USA}
 \author{A~Romero-Wolf}\affiliation{NASA Jet Propulsion Laboratory, California Institute of Technology, Pasadena, CA 91109 USA}
 \author{J~Ziemer}\affiliation{NASA Jet Propulsion Laboratory, California Institute of Technology, Pasadena, CA 91109 USA}



\begin{abstract}
The zodiacal dust complex, a population of dust and small particles that pervades the solar system, provides important insight into the formation and dynamics of planets, comets, asteroids, and other bodies. We present a new set of data obtained from direct measurements of momentum transfer to a spacecraft from individual particle impacts. This technique is made possible by the extreme precision of the instruments flown on the \emph{LISA Pathfinder} spacecraft, a technology demonstrator for a future space-based gravitational wave observatory. Pathfinder employed a technique known as drag-free control that achieved rejection of external disturbances, including particle impacts, using a micropropulsion system. Using a simple model of the impacts and knowledge of the control system, we show that it is possible to detect impacts and measure properties such as the transferred momentum, direction of travel, and location of impact on the spacecraft. In this paper, we present the results of a systematic search for impacts during 4348 hr of Pathfinder data. We report a total of 54 candidates with transferred momenta ranging from 0.2 to 230$\,\mu$Ns.  We furthermore make a comparison of these candidates with models of micrometeoroid populations in the inner solar system, including those resulting from Jupiter-family comets (JFCs), Oort-cloud comets, Hailey-type comets, and Asteroids. We find that our measured population is consistent with a population dominated by JFCs with some evidence for a smaller contribution from Hailey-type comets, in agreement with consensus models of the zodiacal dust complex in the momentum range sampled by \emph{LISA Pathfinder}.
\end{abstract}

\keywords{instrumentation: miscellaneous --- planetary systems: meteoroids}



\section{Introduction} \label{sec:intro}
Our solar system hosts a population of dust and small particles that originate as debris from asteroids, comets, and other bodies.  Understanding these particles is important both for gaining insight into both the formation of our Sun and its planets and the dust population around other stars. More practically, dust and micrometeoroids are a critical component of the environment in which our spacecraft operate and against whose hazards they must be designed.  The behavior of the solar system dust complex has been addressed from both theoretical and observational perspectives. Theorists have developed models of the production of dust from comets and asteroids,  its evolution under the effects of gravity and the solar environment, and its destruction through accretion and other processes. Observationally, this population has been constrained through measurements of its interaction with Earth's atmosphere (photographic, visual, and radio meteors; e.g. \cite{Halliday1984, Hawkes2007, Trigo-Rodriguez2008}), observations of zodiacal light (e.g.,  \cite{Krick2012, Durmont1980}), analysis of microcraters in Apollo lunar samples (e.g.  \cite{Allison1982}), and in-situ measurements made with ionization and penetration detectors on spacecraft (e.g.,  \cite{Weiden1978, Zhang1995}).  These theoretical and observational models are broadly consistent with one another, although important questions remain. One issue is that the bulk of the observational data is from the environment near Earth, a region in which some of the more subtle differences in the models of the underlying population are masked by the influence of the planet itself. Data taken far from Earth could in principle be used to distinguish such subtleties.
\\
\emph{LISA Pathfinder} (LPF, \cite{Antonucci_2011a}), a European Space Agency (ESA) mission that operated near the first Sun--Earth Lagrange point (L1) from 2016 January through 2017 July, is in an ideal orbit to make such measurements. However, \emph{LPF} flew no instrumentation dedicated to micrometeoroid or dust detection.  \emph{LPF}'s primary objective was to demonstrate technologies for a future space-based observatory of millihertz-band gravitational waves. The key achievement of \emph{LPF} was placing two gold-platinum cubes known as `test masses' into a freefall so pure that it was characterized by accelerations at the femto-g level (e.g., \cite{LPF_PRL_2016, LPF_PRL_2018}), the level required to detect the minute disturbances caused by passing gravitational waves. In order to reach this level of performance, the test masses were released into cavities inside the spacecraft and a control system was employed to keep the spacecraft centered on the test masses.  This control system was designed to counteract disturbances on the spacecraft, including those caused by impacts from micrometeoroids. Shortly before \emph{LPF}'s launch, it was realized that data from the control system, if properly calibrated, could be used to detect and characterize these impacts and infer information about the impacting particles (e.g., \cite{Thorpe:2015cxa}). While such impact events have been reported by other spacecraft, \emph{LPF}'s unique instrumentation makes it sensitive to much smaller and much more numerous impacts and allows the impact geometry to be more fully constrained. Early results from the first few months of \emph{LPF} operations suggested that such events could indeed be identified and were roughly consistent with the pre-launch predictions of their effect on the control system (e.g., \cite{Thorpe2017a}). In this paper we present results from the first systematic search for micrometeoroid impacts in the \emph{LPF} data set.  Our data set consists of \nhours hours of data in both the nominal \emph{LPF} configuration as well as the ``Disturbance Reduction System" (DRS) configuration, in which a NASA-supplied controller and thruster system took over control of the spacecraft \citep{ST7_Results}. Our data set corresponds to the times when \emph{LPF} was operating in a `quiet' mode, without any intentional signal injections or other disturbances. During this period, we have identified \nhits impact candidates using our detection pipeline and manual vetoing. We have characterized the properties of this data set and compared it to several theoretical models for the underlying dust population. 
\\
The remainder of the paper is organized as follows. In Section \ref{sec:models} we summarize the dust population models to which we compare our data set and their relevant properties. In Section \ref{sec:methods} we describe our detection technique, including initial calibration, search, parameter estimation, and vetoing. Section \ref{sec:results} summarizes our results, including examples of individual events and properties of the observed population. In Section \ref{sec:model_inference} we present a statistical comparison of our observed population with the theoretical models for the dust population. Conclusions from this work and implications for future work are contained in Section \ref{sec:conclusions}. A complete list of the impact candidates is included in Appendix A.

\section{Population Models}\label{sec:models}
In this work we utilized dynamical models of meteoroids in the solar system to characterize the direction, velocity, and mass of particles impacting the \emph{LISA Pathfinder} spacecraft.
The meteoroids considered here originate from three cometary sources: short-period Jupiter-Family comets (JFCs) and long-period Halley-Type and Oort Cloud comets (HTCs and OCCs, respectively), as well as asteroidal sources (ASTs). JFCs are modeled following the work  reported by~\cite{Nesvorny10,Nesvorny11a}, who estimated that these particles represent 85 $-$ 95\% of the total meteoroid budget (in terms of number of particles) in the inner solar system. The assumed JFCs' initial distribution of orbital elements followed the one proposed by~\cite{LevisonDuncan97}, where the number of comets as a function of their distance from perihelion, $q$, is given by

\begin{equation}
dN(q)\propto q^{\gamma_\mathrm{JFC}} dq\label{nq}
\end{equation}

\noindent where $\gamma_\mathrm{JFC}$ is a free parameter ($\gamma_\mathrm{JFC} = 0$ in this work). The continuous size-frequency distribution (SFD) of meteoroids produced by these comets is given by a broken power law

\begin{equation}
dN(D)\propto D^{-\alpha}dD\label{nd}
\end{equation}

\noindent where $D$ is the meteoroid diameter and $\alpha=4$ the slope index. Once released from the comets, JFC meteoroids drift toward the inner solar system under the influence of Poynting-Robertson (P-R) drag and provide a continuous input of extraterrestrial material to Earth from the direction of the heliocentric and anti-heliocentric apparent sporadic sources~\cite[][]{JonesBrown93,Nesvorny10}.

To describe the contribution of long-period HTCs we utilized the steady-state model by~\cite{Pokorny14}, who used it to explain the origin of the toroidal meteoroid sources~\cite[][]{JonesBrown93,CampbellBrown09,Janches15}, characterized by high ecliptic latitude radiants ($\beta\sim\pm 55^{\circ} - 60^{\circ}$), located both north and south from the apex direction. These meteoroids impact the Earth with a typical velocity of $\sim$35 km~s$^{-1}$, resulting in high-inclination pre-atmospheric orbits with respect to the ecliptic ($\sim70^{\circ}$). In addition, their semimajor axes are close to 1 AU, but with a long tail to larger values, and have a broad distribution of eccentricities with a maximum at $\sim$0.2~\cite[see Figure 13 in][]{Janches15}. 

\begin{figure}[t!]
\gridline{\fig{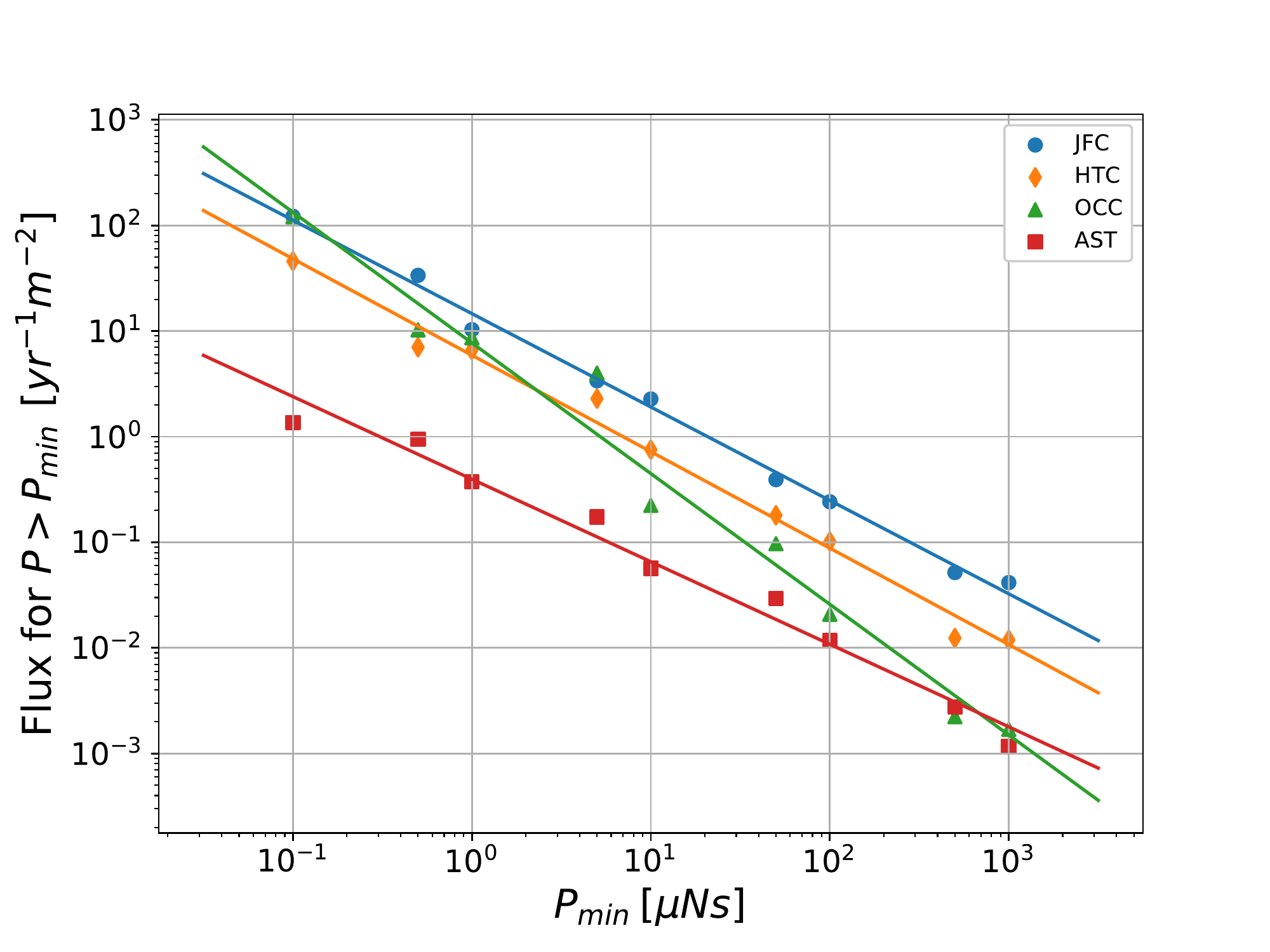}{0.9\columnwidth}{}}
\caption{Expected sky-averaged flux of micrometeoroids in the vicinity of Sun--Earth L1 as a function of momentum relative to L1 and class of parent body. JFC = Jupiter-Family comets, HTC = Halley-Type comet, OCC = Oort Cloud comet, AST = asteroid. See text for details. \label{fig:popCDF}}
\end{figure}

This model tracks the dynamical evolution of thousands of dust particles released from a synthetic population of HTCs for millions of years until particles reach the end of their life, either by being scattered from the solar system by giant planets (mostly Jupiter), or by encountering one of the terrestrial planets, or by evolving too close to the Sun. The model adopts the HTC orbital architecture proposed by~\cite{Levison06} based on an observed inclination distribution of HTCs, which contains preferentially prograde orbits with a median inclination value of $\sim$55$^{\circ}$ and only a small fraction of comets on retrograde orbits. The prograde portion of HTCs populates mostly the toroidal sources with a characteristic velocity distribution that peaks at $\sim$25 km~s$^{-1}$. The model shows also that the aphelion source is formed in part also by HTC-released particles, with a velocity distribution which peaks at $\sim$55 km~s$^{-1}$. These are predominantly retrograde or high-eccentricity orbits representing a minority ($\sim$11\%) of cases among the HTCs, yet, together with OCCs, they probably dominate impact ejecta production at the Moon \citep{Pokorny2019}.

\begin{figure*}[ht!]
\gridline{\fig{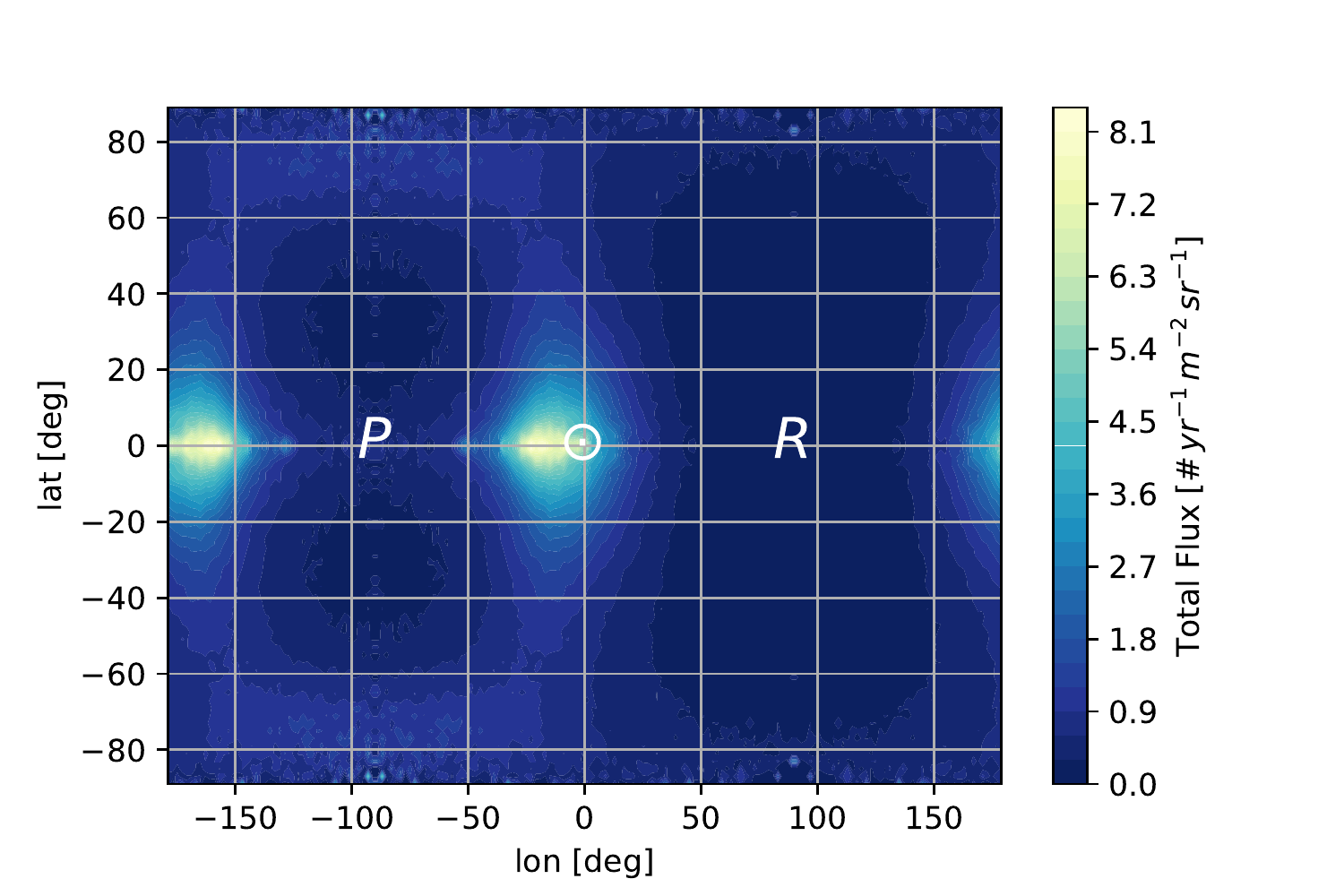}{0.9\columnwidth}{(a) Jupiter-Family Comets (JFC)}
\fig{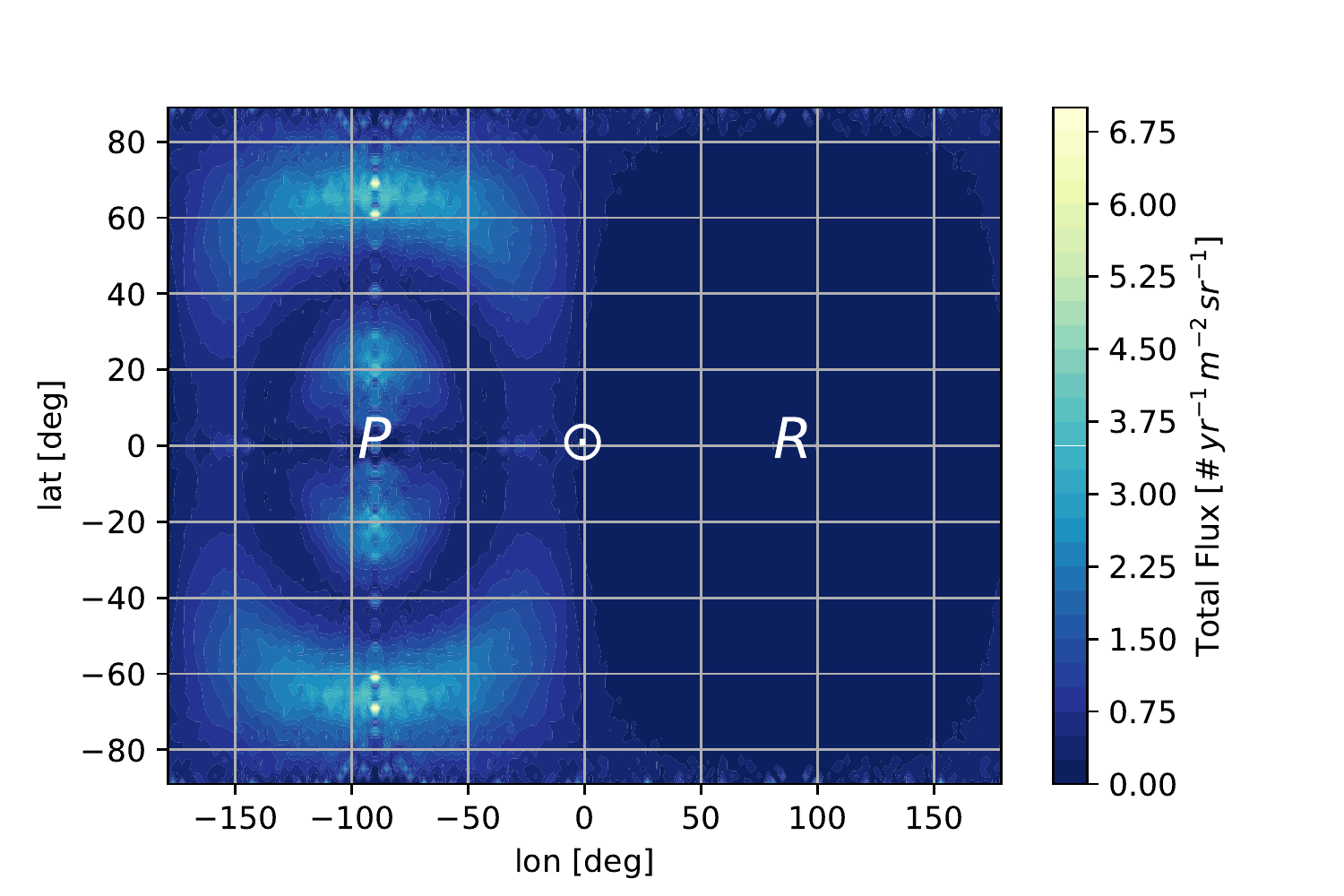}{0.9\columnwidth}{(b) Halley-Type Comets (HTC)}}
\gridline{\fig{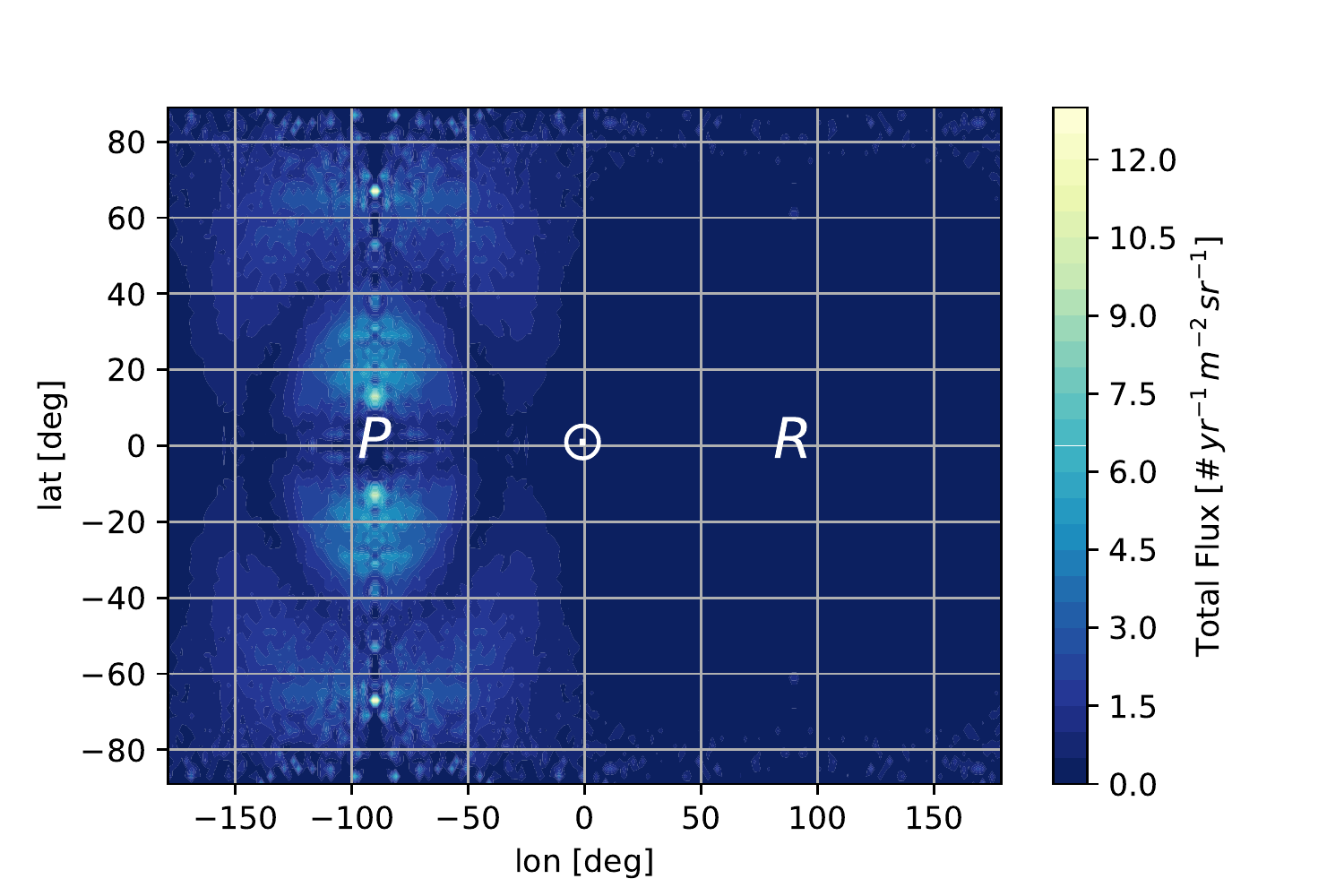}{0.9\columnwidth}{(c) Oort Cloud Comets (OCC)}
\fig{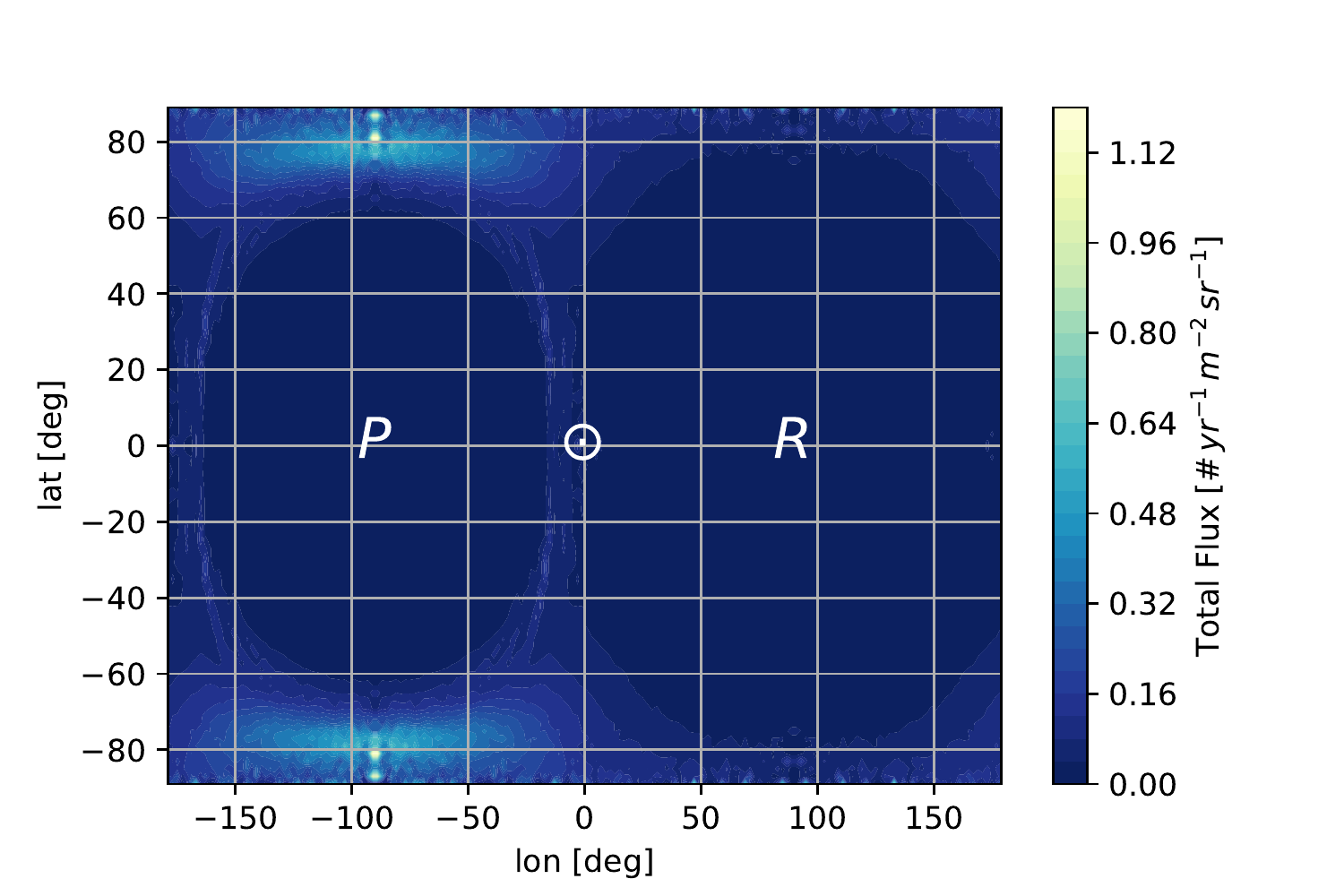}{0.9\columnwidth}{(d) Asteroids (AST)}}
\caption{Angular flux density for micrometeoroid impacts with momenta $\geq 1\,\mu$Ns at Sun--Earth L1 as a function of sky position in a Sun-tracking ecliptic frame with the Sun at 0$^\circ$ longitude (point $\odot$), L1 prograde direction at $-90^\circ$ (point P), and L1 retrograde direction at $+90^\circ$ (point R). 
\label{fig:popMaps}}
\end{figure*}

For meteoroids released from OCCs, we adapted the model developed by~\cite{Nesvorny11b}, who investigated the effects of radiation pressure on particles released from the highly eccentric OCC orbits and their dynamical evolution under gravitational perturbations from planets and P-R drag to determine whether at least a fraction of the near-Earth meteoroid environment is produced by the contribution of dust released from these bodies. 
For small perihelion distances $q$, the model follows the orbital distribution reported by~\cite{Francis05}. For larger perihelion distances, the authors assumed an increasing distribution with $q$, as opposed to the flat and/or declining distribution proposed by~\cite{Francis05}, given by

\begin{equation}
dN(q)\propto
\begin{cases}
(1+\sqrt q)dq & \text{if}\ q<2 \mathrm{AU}\\
2.41(q)^{\gamma_\mathrm{OCC}} dq & \text{if}\ q> 2 \mathrm{AU}\\
\end{cases}
\end{equation}

\noindent where 0$\leq\gamma_\mathrm{OCC}\leq$1 and $q$ is uniformly distributed between 0 AU$\leq q\leq$ 5 AU, thus assuming that particles with $q>$ 5 AU will never reach Earth. In this work, we use $\gamma_\mathrm{OCC}= 0$ since the authors found that the effect of changing this parameter was insignificant.  

\cite{Nesvorny11b} found that OCC particles cannot provide a significant contribution to the overall meteoroid budget of the inner zodiacal cloud. Most of the small particles (i.e. $D\sim$10 $\mu$m) are blown out of the solar system by radiation pressure, while millimeter-sized meteoroids get scattered by planets and their orbits never decouple from Jupiter, and thus the collision probability of these meteoroids with Earth is negligible. 
The authors concluded that only meteoroids with diameters between $\sim$100 and 300 $\mu$m can evolve in orbits decoupled from Jupiter and effectively populate the aphelion source with preferentially retrograde meteors observed impacting the Earth with speeds of around $55-60$ km~s$^{-1}$.  

Micrometeoroids from ASTs are modeled following \cite{Nesvorny10}.

Collectively these models have been utilized to model various meteoroid-related phenomena at Earth~\cite[][]{CarrilloSanchez16, Janches17}, at the Moon, and at Mercury~\citep{Pokorny17,Pokorny18}.  In this paper we use the models to estimate the expected flux for \emph{LPF}'s position at Sun--Earth L1. Figure \ref{fig:popCDF} shows the sky-averaged flux of potential impactors at L1 as a function of the minimum momentum relative to L1, extending down to a momentum of 0.1$\,\mu$Ns, which is the approximate sensitivity limit of \emph{LPF} derived in \cite{Thorpe:2015cxa}. The filled points show outputs of Monte Carlo simulations for particles with parent bodies of the following types: JFCs, HTCs, OCCs, and ASTs.  The numerical results are reasonably well fit by a simple power law in momentum (solid lines) as is commonly used in phenomenological models of micrometeoroid flux (e.g. \cite{Grun1985}). Table \ref{tab:popCDFfits} lists the best-fit parameters and 1$\sigma$ errors for such a fit to each population. Based on these fits, we would expect that events in the \emph{LPF} detection range to be a mixture of roughly 2/3 JFCs, 1/3 HTCs, and a smaller contribution from OCCs and ASTs.  For lower momenta closer to detection threshold, the contribution of OCC events increases, eventually equaling that of the JFCs for a minimum momentum of $0.1\,\mu$Ns.  

\begin{table}[hb!]
\caption{Best-fit parameters and 1$\sigma$ errors for power-law fits to micrometeoroid Monte-Carlo models in Figure \ref{fig:popCDF} of the form $R\,\left(\frac{P_{min}}{1\,\mu\textrm{N s}}\right)^{-\alpha}$. \label{tab:popCDFfits}}
\begin{tabular}{|c|c|c|}
\tableline
population & $R\,\frac{\#}{\textrm{yr}\textrm{m}^2}$ & $\alpha$ \\
\tableline
JFC &  $14^{+6}_{-4}$ & $0.88\pm 0.03$ \\ 
HTC &  $6^{+4}_{-2}$ & $0.91\pm 0.04$ \\ 
OCC &  $8^{+14}_{-5}$ & $1.24\pm 0.09$ \\ 
AST &  $0.4^{+0.3}_{-0.2}$ & $0.78\pm 0.05$ \\ 
\tableline
\end{tabular}
\end{table}

A distinguishing feature of these four populations of micrometeoroids is their sky distribution.  Figure \ref{fig:popMaps} shows maps of angular flux density for micrometeoroid impacts with momenta $\geq 1\,\mu$Ns as a function of sky position in a Sun-tracking Ecliptic Frame centered on L1.  The Sun is located at 0$^\circ$ longitude (point $\odot$), the prograde direction at $-90^\circ$ (point P), and the retrograde direction at $+90^\circ$ (point R).  JFC particles are concentrated into two clumps near the ecliptic plane, one from a roughly Sunward direction and one from a roughly anti-Sun direction. The longitudes of both clumps are shifted slightly towards the prograde direction due to the orbital motion of L1. HTC particles are centered in the prograde direction and distributed in two symmetric sets of clumps above and below the ecliptic plane with median latitudes of roughly $\pm 20^\circ$ and $\pm 65^\circ$.  OCC have a similar distribution to HTCs, although the lower-latitude clumps are more pronounced and slightly closer to the ecliptic. ASTs are concentrated mostly at high latitudes and in the prograde direction. 

Overall, these models predict a detectable impact rate on the order of $10^2$ events per year for the \emph{LPF} spacecraft.  A strong bias towards lower momenta is expected, which predicts that the number of events that are measured well enough to infer sky positions (see discussion in section \ref{sec:sensitivity}) should be considerably smaller. 

\section{Methods} \label{sec:methods}
The process of extracting micrometeoroid impact events from the \emph{LPF} data stream can be divided into three distinct steps: calibration to equivalent free-body acceleration, detection and parameter estimation, and post-processing.  The following three subsections describe these three steps in more detail, the end result of which is a catalog of impact candidates.

\subsection{Calibration of \emph{LPF} data}\label{sec:calibration}
As mentioned in the introduction, \emph{LPF} uses a sophisticated control system to maintain the positions and attitudes of the spacecraft (S/C) and the two test masses (TMs) such that a number of constraints are satisfied.  Example constraints include maintaining the positions and orientations of the TMs at constant values relative to the S/C and maintaining the S/C attitude relative to the Sun and Earth. In total, the control system takes measurements of 15 kinematic degrees of freedom (dofs), 3 positions + 3 attitudes for both test masses and 3 attitudes for the spacecraft, and generates actuation commands for 18 dofs, 3 forces and 3 torques for the two test masses and the spacecraft.  Positions and angles are measured using a star tracker, a capacitive sensing system, and an optical interferometric sensing system. An electrostatic actuation system applies forces and torques to each TM, and a micropropulsion system applies forces and torques to the S/C. 

\begin{figure}
\gridline{\fig{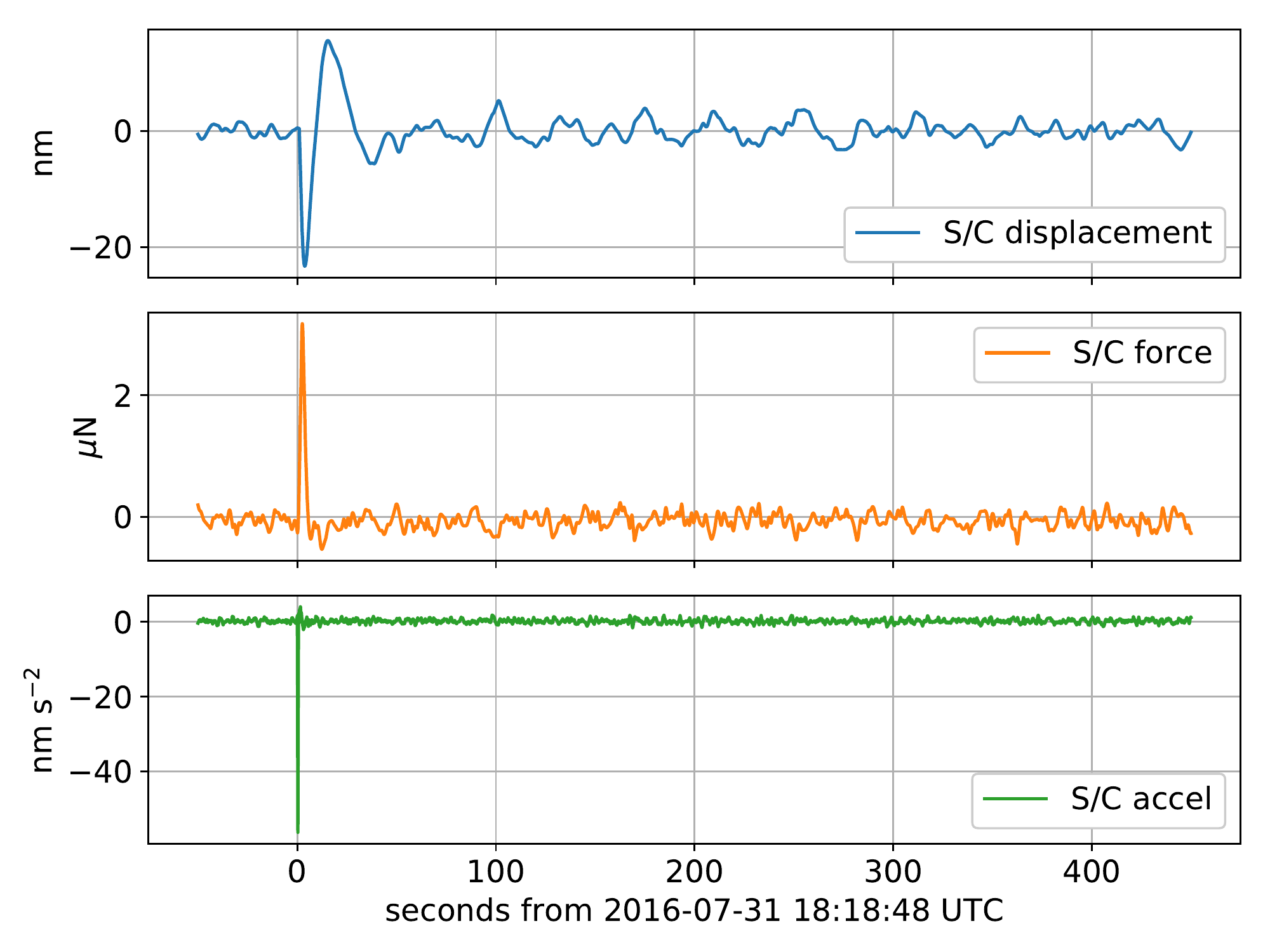}{1.0\columnwidth}{}}
\vspace*{-10mm}
\caption{Example of x-axis telemetry for impact candidate at GPS time 1154024345.4 (2016 July 31 18:18:48 UTC) and the equivalent free-body acceleration estimated through the calibration procedure. The top panel shows the displacement of the S/C in the $x$-direction. The middle panel shows the commanded force on the S/C in the $x$-direction by the control system. The bottom panel shows the reconstructed external acceleration on the S/C in the $x$-direction using the above data and S/C geometry and mass properties. \label{fig:calExample}}
\end{figure}

One effect of the control system is to split the effect of a micrometeoroid impact into both the measured position and commanded force signals, both of which are telemetered to ground. Figure \ref{fig:calExample} shows an example of this for an impact candidate observed on 2016 July 31.  The top panel shows the measured position of one TM relative to the S/C along the x-axis, as measured using the optical interferometer.  For the  $\sim50\,$s prior to the event the signal exhibits random fluctuations with an RMS amplitude of a few nanometers. At the time of the event, the signal shows a steep downward ramp, reaching more than $20\,$nm in a few seconds.  The middle panel of Figure \ref{fig:calExample} shows the force commands on the S/C in the $x$-direction, which are used to maintain the TM-S/C distance in this control mode. Shortly after the observed ramp in the motion, the controller commands a thrust of a few $\mu$N in the $+x$-direction to compensate this motion.  The resulting acceleration of the S/C causes the TM-S/C separation to stop increasing, turn around, and return toward zero. In response, the controller reduces the applied force on the S/C. After two oscillations and roughly a minute, the system is back in its quiescent state.  By combining the force telemetry and the position telemetry with appropriate constants such as the calibration of the force actuators and the mass of the S/C, the equivalent free-body acceleration can be constructed. This is shown in the bottom panel of Figure \ref{fig:calExample} and exhibits the classic impulse response in acceleration that is expected for an impact. 

The basic process illustrated in Figure \ref{fig:calExample} can be repeated along the other dofs of the S/C in order to develop a data set of the equivalent free-body acceleration of the S/C in all 6 dof. In doing so, a number of considerations must be addressed. First, the fact that the TMs are not located at the center of mass of the S/C means that torques applied to the S/C lead to accelerations in the linear dof of the S/C.  Secondly, the `topology' of the control system, or which actuations are used to control which displacements, is different for each dof and also for the various operational modes of the control system. Lastly, generating the free-body accelerations requires knowledge of a number of calibration factors such as S/C and TM mass and moments of inertia, location of the TMs in the S/C frame, locations of the thrusters on the spacecraft and their orientations, calibration and cross-talk in both sensors and actuators, and relative timing/phase information between the various telemetries.  Some of these effects and calibration factors were measured in flight during dedicated experiments designed to calibrate the \emph{LPF} hardware for its primary mission. Examples include calibration of the x-axis electrostatic TM actuation (\cite{LPF_Calibration}) and calibration of the thruster response (\cite{ST7_Results}). For quantities that were not measured in flight, our models were built using the nominal values provided by the equipment manufacturers. 
\\
The end result of this calibration step was a set of 12 time series corresponding to the equivalent free-body acceleration of the S/C in each of 6 dof as measured by each of the two TMs. We denote these as $g_{1i}(t)$ and $g_{2i}(t)$ where $i=\left(x,\,y,\,z,\,\theta,\eta,\,\phi\right)$ for TM1 and TM2, respectively. The S/C coordinates are defined such that $z$ is the direction of the top deck (oriented at the Sun), $x$ is the direction along the two test masses with +x pointing from TM1 toward TM2, and $y$ completes a right-handed triad. The angles $\theta,\,\eta,\,\phi$ represent right-hand rotations around $x,\,y,\,z$ respectively.


\subsection{Impact Model and Sensitivity}
\label{sec:sensitivity}
The characteristic timescales of the impact process are short relative to the sample cadence of the \emph{LPF} data (typically $0.1\,\textrm{s}$). Consequently, we model the impact as a delta-function impulse in acceleration for each dof.  These impulses occur at the same time for each dof but have different amplitudes that encode information about the impact direction and location on the spacecraft. The modeling of the impact is performed in two steps. First, the acceleration in the S/C body frame is computed for both linear and angular dof:
\begin{eqnarray}
\vec{a}_{x,B}(t) = P\:M^{-1}\delta(t-\tau) \hat{e},\label{eq:axB} \\
\vec{a}_{\theta}(t)= P\:\mathbf{I}^{-1}\delta(t-\tau)\left(\vec{r}\times\hat{e}\right), \label{eq:aq}
\end{eqnarray}
where $\vec{a}_{x,B}$ is the acceleration of the spacecraft body frame in the linear dof, $\vec{a}_{\theta,B}$ is the acceleration of the spacecraft body frame in the angular dof, $P$ is the total transferred momentum, $\tau$ is the impact time, $\hat{e}$ is the unit-vector in the direction of the transferred momentum, $M$ is the mass of the S/C, $\mathbf{I}$ is the S/C moment of inertia about its center of mass, and $\vec{r}$ is the location of the impact relative to the center of mass. The angular accelerations at the TM locations are the same as described in Equation (\ref{eq:aq}), but the linear accelerations pick up an additional term due to the offset of the test mass from the center of mass:
\begin{equation}
\vec{a}_{x,TM}(t) = \vec{a}_{x,B} + \left(\vec{r}_{TM}\times \vec{a}_{\theta}\right),\label{eq:axTM}
\end{equation}
where $\vec{a}_{x,TM}$ is the acceleration in the linear dof as measured in the test mass frame and $\vec{r}_{TM}$ is the location of the test mass relative to the S/C center of mass.

Sensitivity to impacts is limited by two noise sources: noise in the measurement system and disturbances on the S/C.  Measurement noise for both the capacitive and interferometric systems is characterized by a white spectrum in displacement, whereas the chief noise source for the S/C disturbance, the micropropulsion system itself, exhibits an approximately white spectrum in force.  The relative levels of these two components differ for each dof,, but the basic functional form for the noise power spectral density is
\begin{equation}
S_{g}=S_0+S_4f^4,
\label{eq:noise}
\end{equation}
where $S_0$ is the amplitude of the S/C disturbance term and $S_4$ is the amplitude of the measurement term. The most substantial difference between the noise level in the various dof is in the amplitude of the $S_4$ term, which is substantially lower for the dof sensed by the interferometric system: $x,\,\eta,$ and $\phi$.  In \cite{Thorpe:2015cxa} it was shown that signal-to-noise ration (S/N) of a simple impulse in the presence of this noise shape can be analytically computed as $\rho =  P_i/P_c$, where $P_i$ is the amplitude of the momentum transfer in that dof and $P_c$ is a characteristic threshold momentum given by
\begin{equation}
P_c \equiv \frac{1}{\sqrt{2\pi}}\left(4 S_4 S_0^3\right)^{1/8}.\label{eq:SNRp}
\end{equation} 
The value of $P_c$ varies somewhat for each dof owing to the different combinations of sensing noise and micropropulsion noise, as well as differences in the spacecraft mass properties. The approximate range is $0.05\,\mu\textrm{N\:s}\leq P_c \leq 1\,\mu\textrm{N\:s}$ for linear dof and $0.3\,\mu\textrm{N\:m\:s}\leq P_c \leq 4\,\mu\textrm{N\:m\:s}$ for angular dof. This asymmetry in sensitivity along different dof means that impacts with lower overall momentum are often only detected in a fraction of dof channels, meaning that the full set of parameters cannot be extracted. Similarly, impacts that happen to impart a large fraction of their momentum in a sensitive channel may be measured at lower thresholds than those coming from different directions.

An important feature of Equation (\ref{eq:aq}) is that the momentum $P$ represents the momentum that is \emph{transferred} to the spacecraft, which may differ from the \emph{intrinsic} momentum of the impacting particle, $\Gamma$. The relationship between these two momenta is typically represented by an impact coefficient, $\beta$, where $P=\beta\cdot\Gamma$.  The value of $\beta$ is highly dependent on the detailed physics of the impact, including the impact geometry, impact velocity, and material properties of both impactor and target~\citep{2018P&SS..164...91F}. For this study we assume a median value of $\beta=3$ and a range $1\leq\beta\leq 5$.  We additionally assume that the value of $\beta$ is independent of impact direction (different impact geometry), impact location (different spacecraft materials), and impact (potentially different impactor composition). While it would in principle be possible to incorporate detailed models of these effects into our analysis, this is beyond the scope of this paper.  Instead, we report both the measured transferred momentum and estimated intrinsic momentum, including errors associated with uncertainty in $\beta$. We do not include additional uncertainty in impact direction, $\hat{e}$, or impact location, $\vec{r}$, associated with a potential dependence on $\beta$ as a function of angle of impact incidence.

\subsection{Detection and Parameter Estimation}\label{sec:MCMC}
The second step in our micrometeoroid pipeline involves the identification and characterization of candidate events in our data stream.  This is performed using the template-matching formalism that is commonly applied in gravitational wave data analysis. 
Assuming that the frequency domain data $\data$ contain an impact signal $\model$ plus noise $\n$, and $\n$ is zero-mean Gaussian distributed, the likelihood for observing  $\data$ is
\begin{equation}
p(\data|\params) = \prod_f \frac{1}{\det \Cij} e^{-\frac{1}{2}\sum_{ij} \residual_i \invCij \residual_j }
\label{eq:likelihood}
\end{equation}
where the $\residual = \data - \model(\params)$ is the residual, $\model(\params)$ is the modeled \emph{LPF} response to an impact with parameters $\params$, and  $\Cij \equiv \langle \n_i \n_j\rangle$ is the one-sided noise correlation matrix. The indices $i$ and $j$ sum over different data channels, i.e. the 6 dof $i:=(x,y,z,\theta,\eta,\phi)$.

We make the simplifying assumption that the noise correlation matrix $\Cij$ is diagonal, i.e. that the noise in each channel is independent. While this is likely a reasonable assumption for the sensing noise component, the platform noise may be somewhat correlated owing to common contributions from the micropropulsion system. 
We further assume that the noise in each channel is stationary, implying that there are no  correlations between different frequencies, and the noise is completely characterized by its variance
\begin{equation}
\langle \n^2(f)\rangle \equiv \frac{T}{2}S_n(f)
\end{equation}
where $T$ is the duration of the data segment and $S_n(f)$ is the one-sided noise power spectral density.
For flexibility to fit realistic instrument noise, we use a phenomenological model for $S_n(f)$ rather than the theoretical form in Eq.~\ref{eq:noise}.
The model is adopted from the BayesLine algorithm~\citep{Littenberg_15} used for spectral estimation in analysis of transient sources detected by the ground-based gravitational wave detector network.
BayesLine is a trans-dimensional (or reversible jump) Markov chain Monte Carlo (MCMC) algorithm~\citep{Green_95}.
The spectral noise model is built from two components:  the broadband spectral shape is fit with a cubic spline interpolation, where the number and location of spline control points are free parameters, and a linear combination of  Lorentzians to fit narrowband spectral lines that were present when the cold gas micropropulsion system was active \citep{ST7_Results}.
The model is flexible and proved to be well suited for fitting the \emph{LPF} noise.

The signal model was implemented as described in Sec.~\ref{sec:sensitivity}, again using a trans-dimensional MCMC.
The MCMC samples between hypotheses that the data contain only noise (i.e. that there is no impact signal in the model) and that the data contain noise and a single impact.
The ratio of MCMC iterations spent in the two hypotheses is the Bayes factor, or marginalized likelihood ratio, between the signal and the noise model.
We use the Bayes factor $B_{\rm signal,noise}$ as the detection statistic, with a threshold of $B_{\rm signal,noise} > 3{:}1$ for claiming a positive detection.
The Markov chain's samples from iterations that included the signal hypothesis are used to characterize the posterior distribution function of the impact parameters, conditional on a signal actually being present in the data.
Marginalized posterior distributions for the incident direction of the impact, as well as the momentum imparted to the spacecraft, are used to make further inferences about the micrometeorite population.
The priors for the signal parameters are uniform distributions in time, imparted momentum, impact location, and incident direction of the impact.  

Both the noise model and signal model MCMC samplers use parallel tempering to improve the convergence time of the chains.
The MCMC code went through a standard suite of tests to confirm that the results are accurate and robust.
The spectral estimation code is validated by testing that the whitened data $\data(f)/\sqrt{S_n(f)}$ are consistent with being drawn from a zero-mean, unit-variance Gaussian.
We check detailed balance of the sampler by using a constant likelihood function and testing that the recovered distributions are consistent with the priors.
Finally, the samplers are tested for accuracy by analyzing simulated and real data with artificial signals added, verifying that the true signal parameters are included in the posterior distributions.
 
The noise and impact models used in this analysis are not perfect, and further advancements may improve the detection efficiency and/or reduce systematic errors in parameter recovery.
A particular weakness is our assumption that the noise in each channel is independent. 
The sensing dof are not the same as the kinematic dof, so noise correlations are not necessarily negligible. 
We also found that, for large momentum impacts, a noticeable residual was left in the data, indicating that our signal model was not a perfect match to the data. 
This modeling mismatch results in an uncharacterized systematic error, though the macroscopic conclusions drawn from the posterior--which face of the spacecraft was impacted, from what (general) region of the sky did the impactor originate, and the overall distribution of imparted momenta of the impactors--are not expected to be biased to the point of misleading the general conclusions.
Improvements to the model, especially developing a physically motivated forward model of the instrument noise, are areas for future study.

\begin{figure*}[t]
\includegraphics[width=\textwidth]{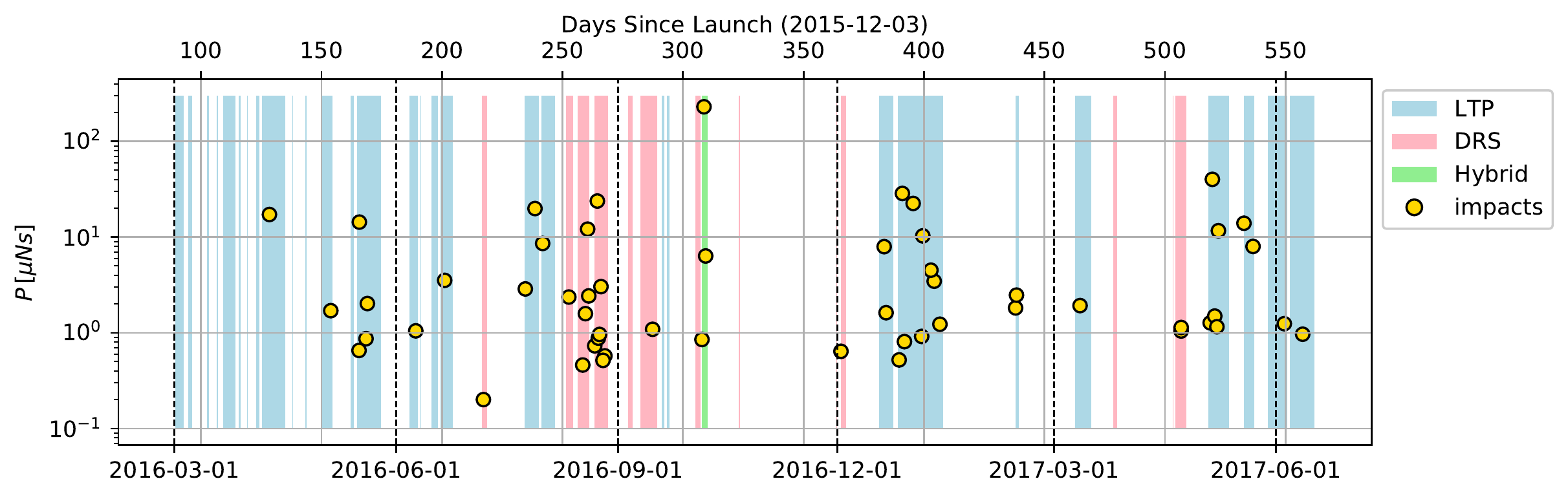} 
\caption{Timeline of impact events during \emph{LPF}. The yellow circles show the impact times, with the total transferred momentum defining the vertical axis. The vertical bars denote the times included in the search, with blue representing the nominal LTP configuration, pink the DRS configuration, and green the hybrid configuration. See text for details. \label{fig:timeline}}
\end{figure*}

\subsection{Post-processing and vetos}\label{sec:vetos}
For each segment of data, the MCMC tool described in Section \ref{sec:MCMC} was run for an initial search composed of $7\times10^4$ steps on the TM1 data. After discarding the first $3\times10^4$ steps of the chain as ``burn-in'' samples, the detection fraction was computed as the ratio of chain steps where an impact model was included to the total number of steps.  For systems with a detection fraction above 0.5, the MCMC tool was rerun in a characterization step of $7\times10^5$ steps on both TM1 and TM2. A burn-in period of $3.5\times10^5$ steps was discarded from both chains, and the detection fraction was again computed, as well as the variance in the impact time parameter $\tau$.  Systems with an above-threshold detection fraction and an impact time variance of less than 0.3\,s in both TMs were passed on to the next step in the vetting process -- manual inspection.  For the manual inspection process, an expanded set of telemetry from the spacecraft around the candidate impact time was downloaded and examined.  Examples of signals inspected include all force and torque signals, all position and attitude signals, selected voltage levels, and internal telemetry of the micropropulsion system. This process yielded two types of false triggers: thruster current spikes and data gaps. Candidates for which the signals appeared consistent with expectations were added to the catalog.

For vetted impact candidates, an additional post-processing step was conducted to extract parameters of interest.  In order to compare with the micrometeoroid population models in Section \ref{sec:models}, it was necessary to transform the impact direction from S/C coordinates to the Sun-tracking ecliptic frame used by the micrometeoroid population models.  This transformation was done in two steps, first from the S/C frame to an Earth-centered inertial (ECI) frame using the S/C quaternion telemetry provided by the star tracker, and then from ECI to the Sun-tracking ecliptic frame using the S/C ephemeris. Median sky location and a 90\% confidence sky area for both frames were computed using HEALPIX\citep{HEALPIX}. 

\begin{figure}[hb!]
\includegraphics[width=\columnwidth]{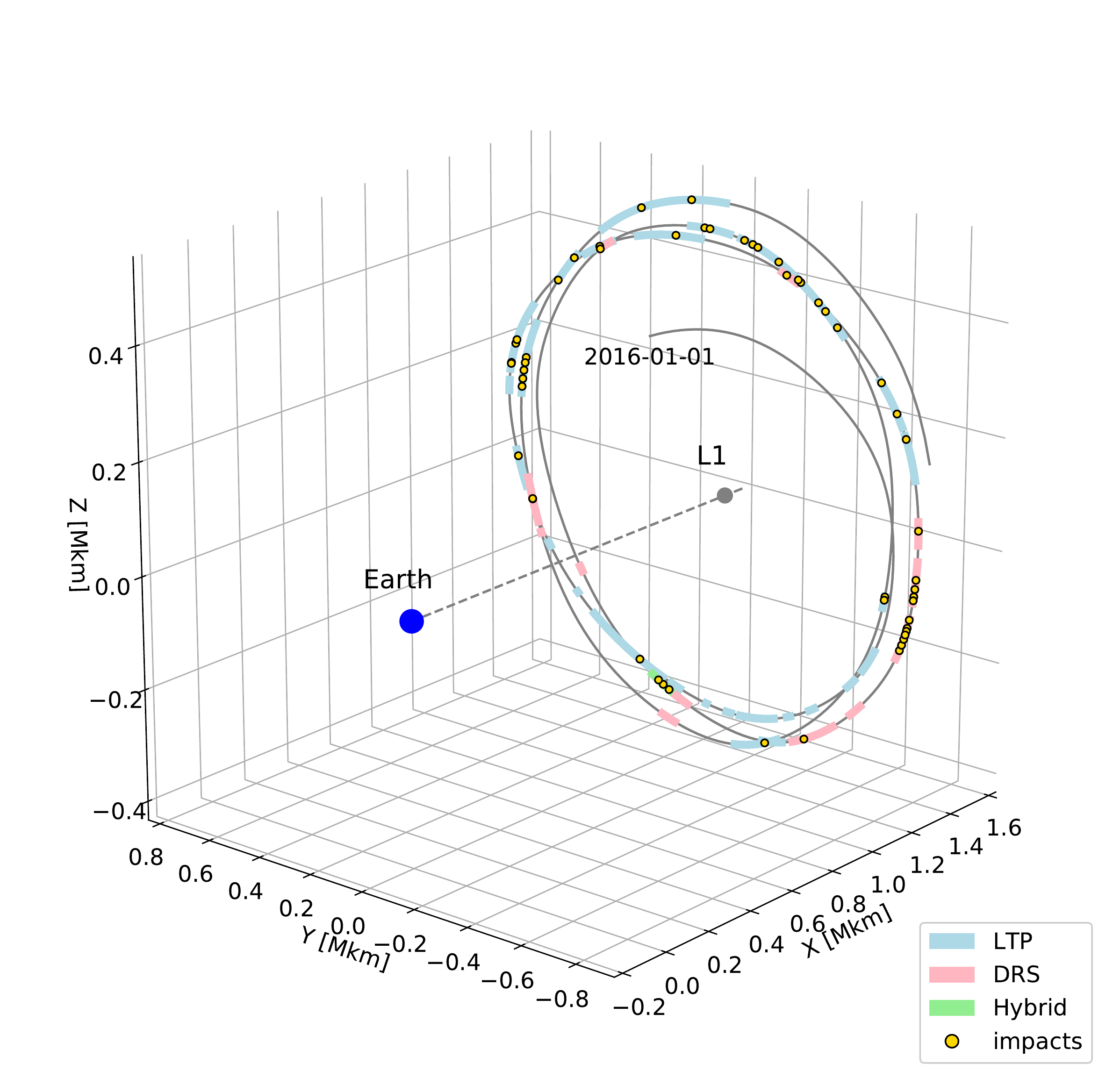} 
\caption{Micrometeoroid impacts visualized along \emph{LPF}'s trajectory as plotted in an Earth-Centered, Sun-synchronous frame.  The solid gray line shows \emph{LPF}'s clockwise trajectory from 2016 January 1 through 2017 July 31 with the times searched for impacts in the LTP, DRS, and hybrid configurations in blue, pink, and green, respectively. Impacts are indicated by yellow circles. \label{fig:ephemeris}}
\end{figure}

\section{Results} \label{sec:results}
In this paper we restrict our analysis to segments of data where no signals were deliberately injected into the \emph{LPF} system. We identified a total of 4348 hr of data in three distinct configurations: the nominal \emph{LPF} configuration, in which the European-provided DFACS control system and cold gas micropropulsion system were operating (3484 hr); the DRS configuration, in which the NASA-provided DCS control system and colloidal micropropulsion system were operating (796 hr); and a hybrid configuration, in which the DFACS was controlling the S/C using the colloidals (61 hr). Figure \ref{fig:timeline} shows a timeline of these segments along with the detected impacts plotted with their total transferred momentum along the vertical axis. The total number of detected impacts is 54: 36 in the nominal configuration, 15 in the DRS configuration, and 3 in the hybrid configuration. This corresponds to a rough event rate of 120 yr$^{-1}$, which is broadly consistent with the estimate made in \cite{Thorpe:2015cxa} as well as the models in Section \ref{sec:models}.  Figure \ref{fig:ephemeris} shows the timeline from Figure \ref{fig:timeline} projected onto the \emph{LPF} ephemeris from 2016 January 1 through 2017 March 31 in an Earth-centered, Sun-synchronous frame.
 
In the following sections we present some example events in detail and summarize some properties of the observed population.  A full catalog of the impacts and their estimated parameters can be found in Appendix A.

\begin{figure*}[]
\gridline{
\fig{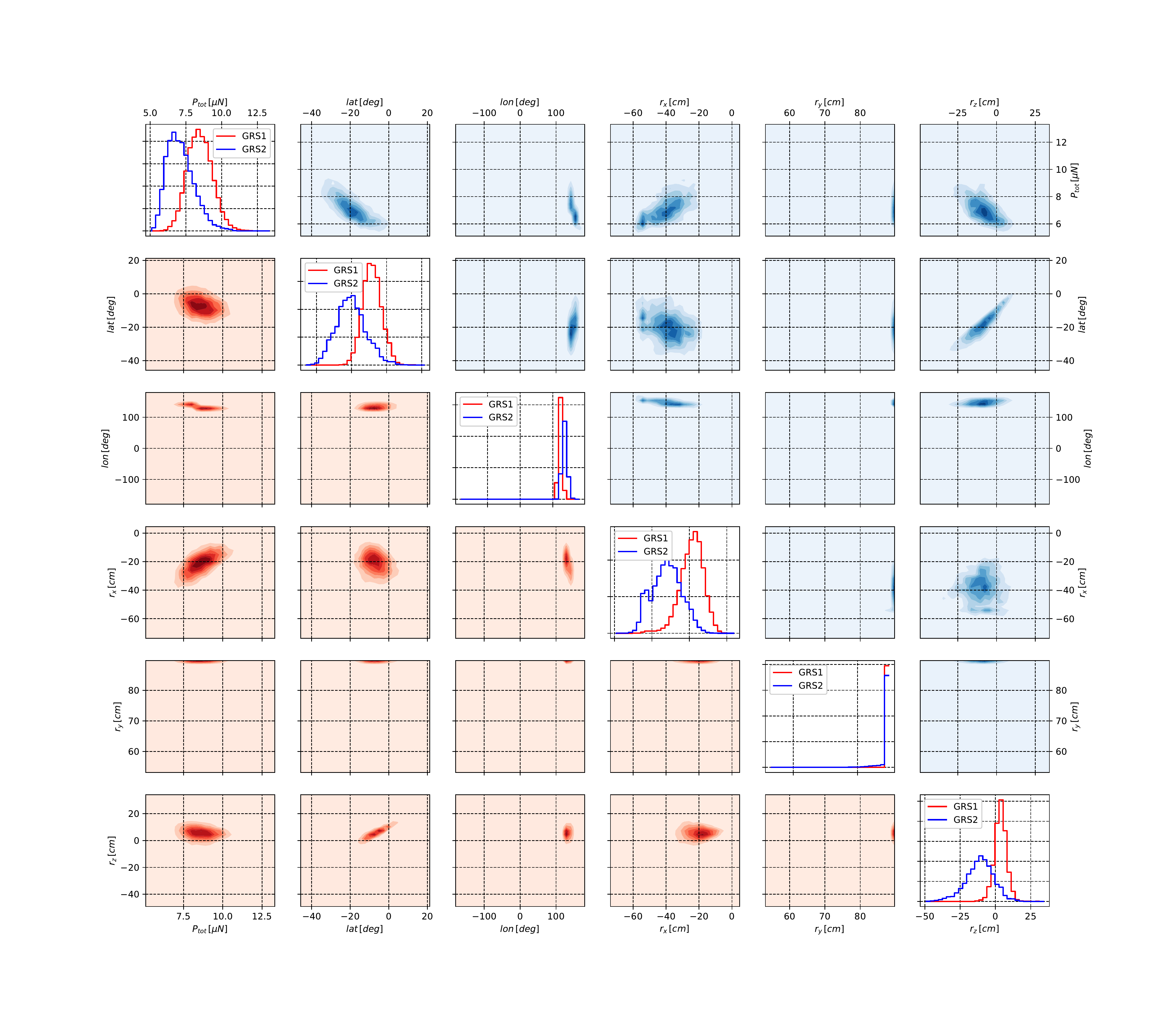}{\textwidth}{}}
\vspace*{-20mm}
\caption{Comparison of recovered posterior distributions for impact parameters using TM1 and TM2 data for the impact candidate occurring at $t_{gps}=1154024345.4$, which is representative of a well-characterized event in our catalog. The array of plots is organized by parameter, with a parameter order from left to right and top to bottom of total momentum transfer, latitude and longitude of impact direction in  spacecraft frame, and $x$, $y$, $z$ location of impact with respect to S/C center of mass. Diagonal panels show single-parameter probability density functions, with TM1 data in red and TM2 data in blue. Lower corner panels (red shades) show two-parameter histograms for TM1, while upper corner panels (blue shades) show two-parameter histograms for TM2. \label{fig:goodExampleCorner}}
\end{figure*}

\begin{figure*}[]
\gridline{
\fig{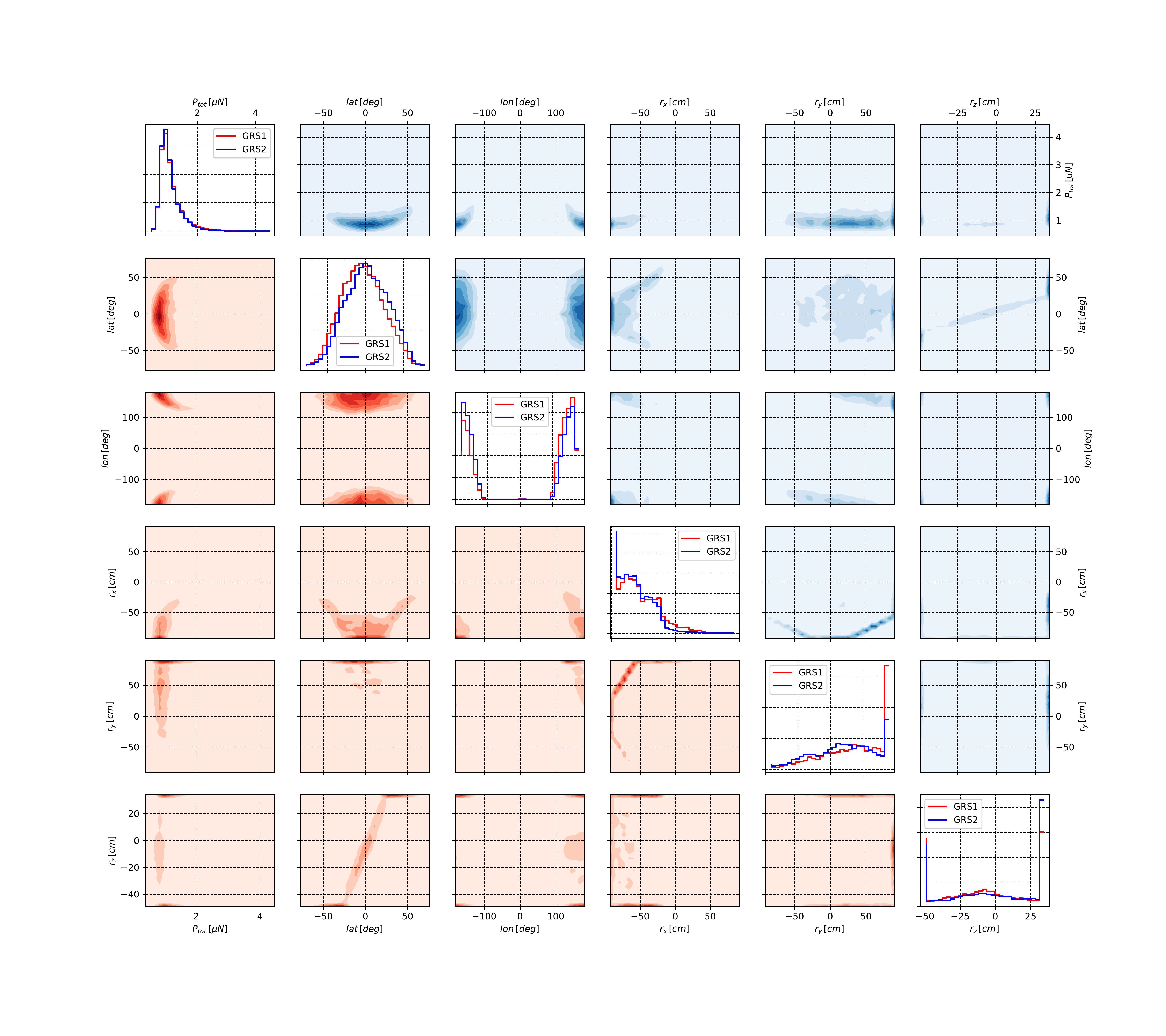}{\textwidth}{}}
\vspace*{-20mm}
\caption{Comparison of recovered posterior distributions for impact parameters using TM1 and TM2 data for the impact candidate occurring at $t_{gps}=1149475987.7$, which is representative of a typically characterized event in our catalog. The plot arrangement is the same as Figure \ref{fig:goodExampleCorner}.\label{fig:typExampleCorner}}
\end{figure*}

\begin{figure*}[h!]
\gridline{
\fig{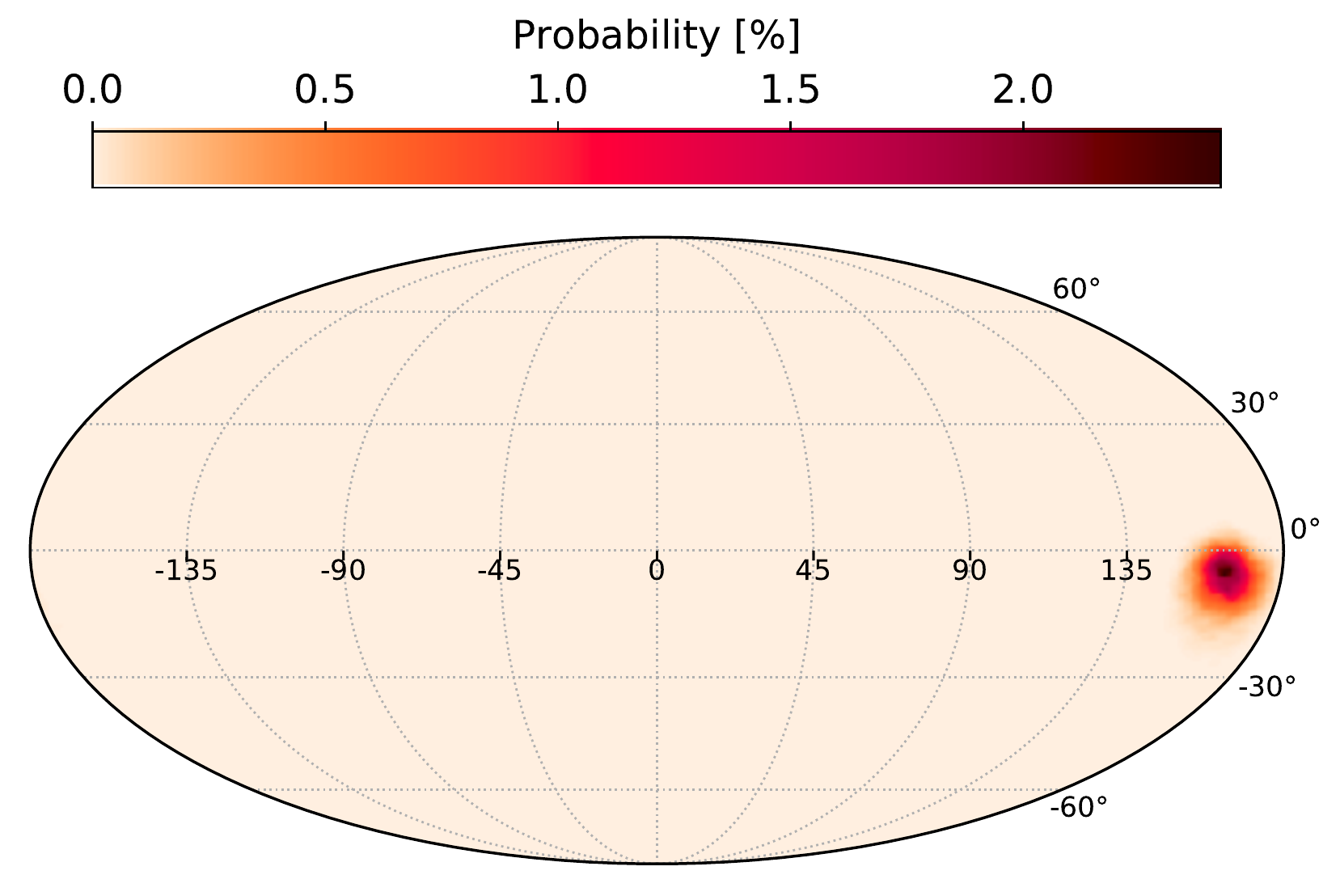}{0.5\textwidth}{(a) Impact origin in Sun-tracking Ecliptic Frame}
\rotatefig{90}{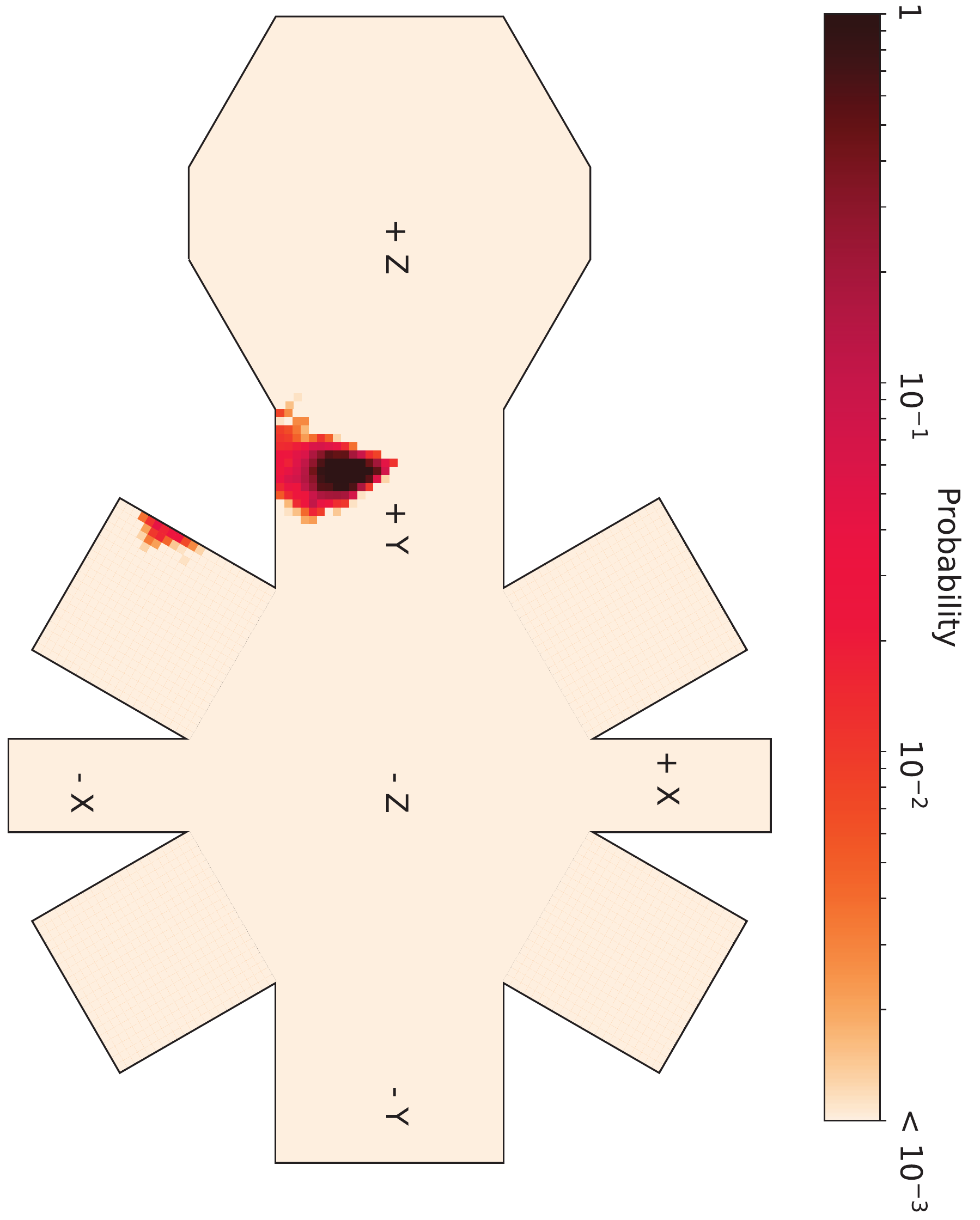}{0.37\textwidth}{(b) Impact location on S/C}}
\caption{Reconstructed impact direction and location using TM1 data for impact candidate occurring at $t_{gps}=1154024345.4$. Color contours denote fraction of post-burn-in MCMC samples in each bin.\label{fig:goodExampleLocation}}
\end{figure*}

\begin{figure*}[h!]
\gridline{
\fig{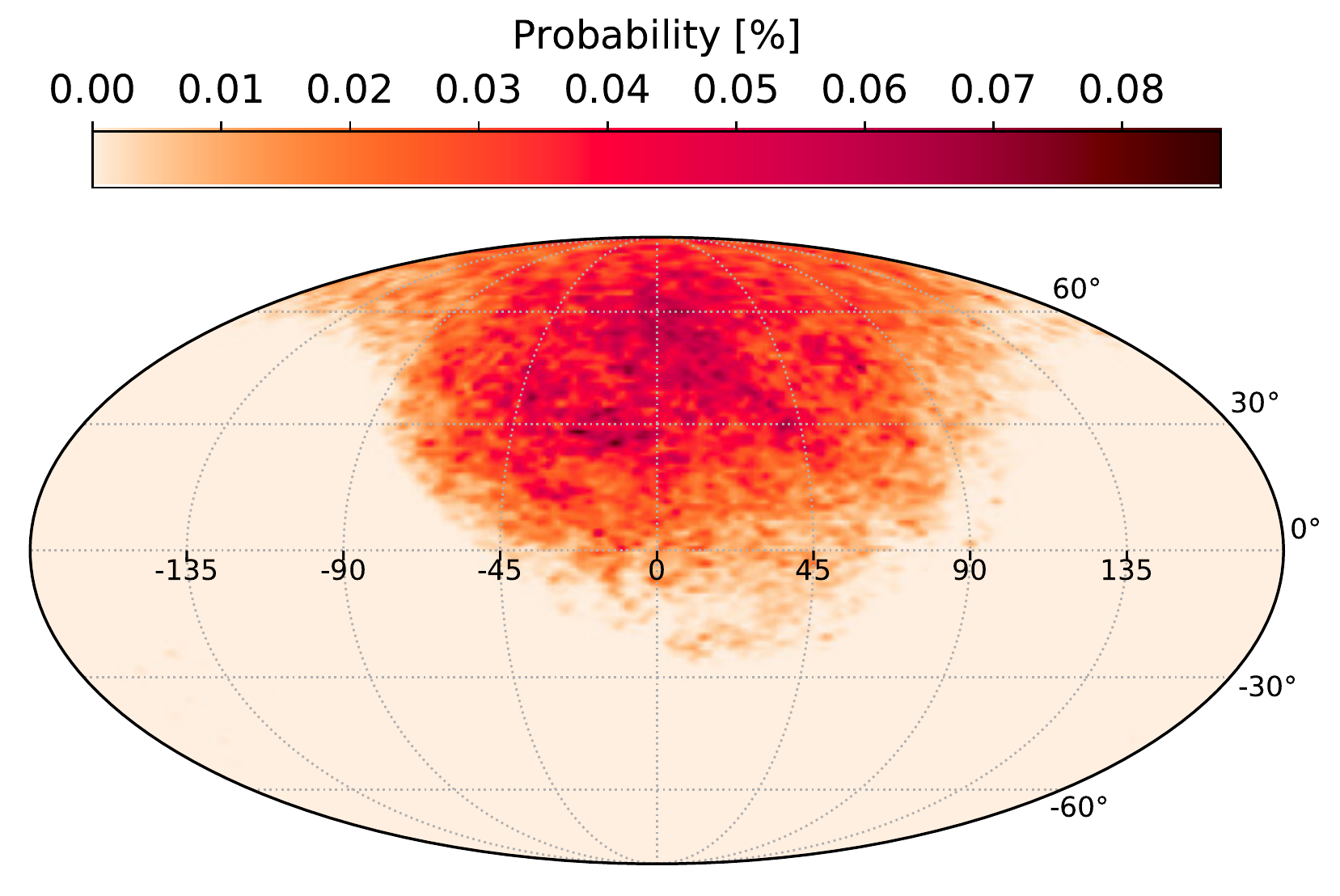}{\columnwidth}{(a) Impact origin in Sun-tracking Ecliptic Frame }
\rotatefig{90}{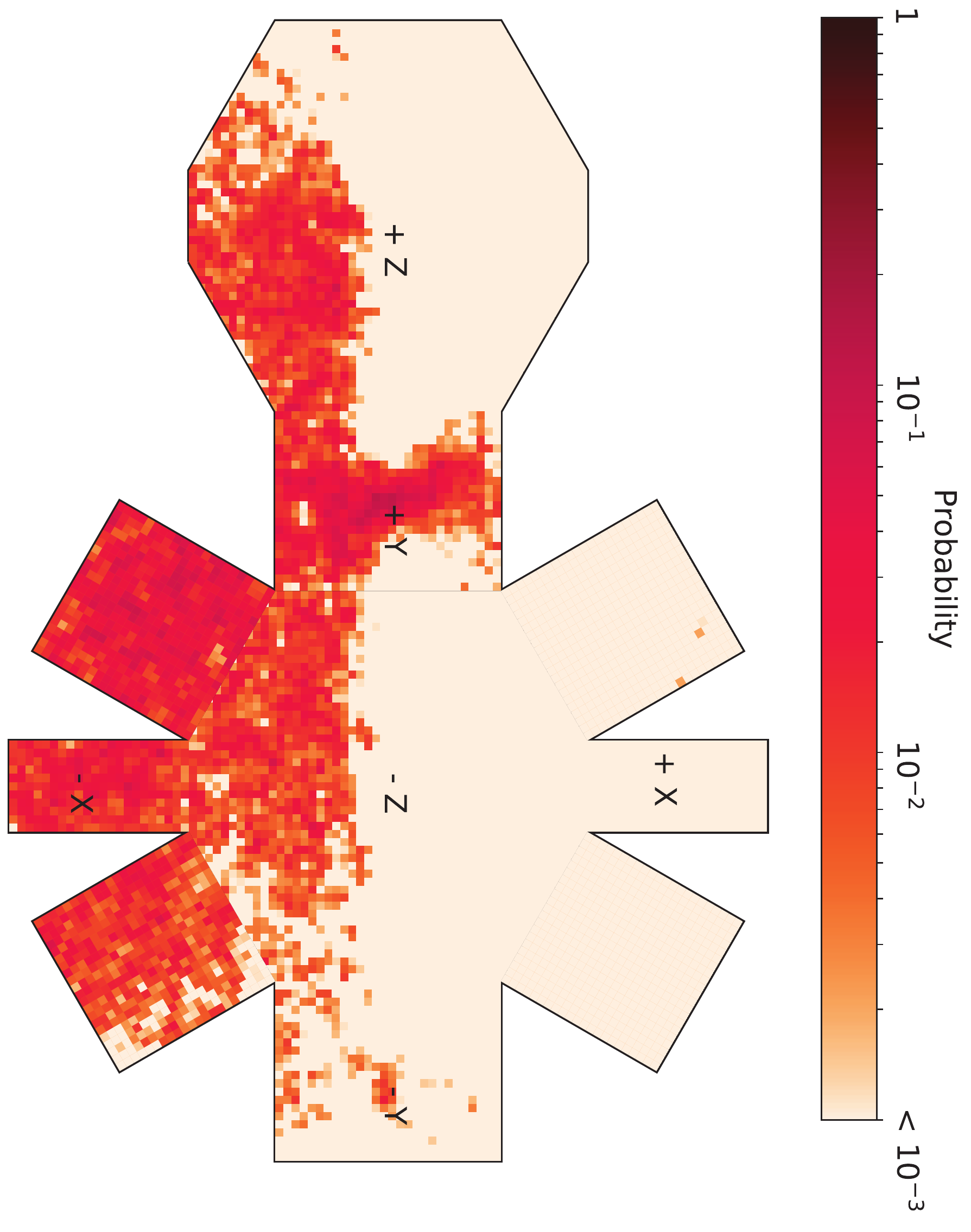}{0.37\textwidth}{(b) Impact location on S/C}}
\caption{Reconstructed impact direction and location using TM1 data for impact candidate occurring at $t_{gps}=1149475987.7$. Color contours denote fraction of post-burn-in MCMC samples in each bin.\label{fig:typExampleLocation}}
\end{figure*}

\subsection{Sample candidate events \label{sec:samples}}
As mentioned in Section \ref{sec:sensitivity}, \emph{LPF}'s sensitivity depends on the parameters of the impact, including both the total momentum transferred and the fraction of that momentum that is  projected into each dof.  As a result, the quality of our parameter estimation varies greatly from impact to impact.  Figures \ref{fig:goodExampleCorner} and \ref{fig:goodExampleLocation} show results for an impact occurring at GPS time $t_{gps}=1154024345.4$, corresponding to 2016 July 31 18:18:48.400 UTC. 

With a moderately high transferred momentum of $8.5\,\mu\textrm{N-s}$ and a S/C longitude that aligns well with the sensitive x-axis, the total S/N's in the two TMs are $\rho_1\approx16$ and $\rho_2\approx22$ . Figure \ref{fig:goodExampleCorner} shows an overlaid corner plot representing the posterior probabilities for the impact parameters as measured by TM1 (in red, lower left) and TM2 (in blue, upper right).  The panels are arranged in a grid, with rows and columns corresponding to the following parameters: total transferred momentum ($P_{tot}$, in $\mu$N-s), S/C latitude defined relative to the S/C $x$--$y$ plane ($lat$, in deg.), S/C longitude defined relative to the $+x$ axis ($lon$, in deg.), and $x$, $y$, $z$, locations of the impact with respect to the S/C center of mass ($r_x,\,r_y,\,r_z$ in $m$).  The panels along the diagonal show the posterior probability density for each parameter as measured by TM1 (red) and TM2 (blue).  The panels on the off-diagonals show the correlation between pairs of parameters in the TM1 data (lower off-diagonals) and TM2 data (upper off-diagonals). The measured parameters between these two impacts are broadly consistent, although TM1 generally prefers a solution with slightly increased $P_{tot}$, larger $lat$, and positive shifts in both $r_x$ and $r_z$. 

Figures \ref{fig:typExampleCorner} and \ref{fig:typExampleLocation} show a similar set of plots to Figures \ref{fig:goodExampleCorner} and \ref{fig:goodExampleLocation} but for an impact occurring at $t_{gps} = 1149475987.7$ (2016 June 9 02:52:50.700 UTC) that had a lower total momentum ($P_{tot}\approx1.0\,\mu\textrm{N-s}$), and lower S/N ($\rho_1\approx1.5$, $\rho_2\approx1.4$). As a result, the constraints on parameters other than the total momentum are rather weak.  The impact location is favored toward the $-x$ and $+z$ faces and the preferred direction to the impactor is in the direction from the Sun (latitudes around 0$^o$) and above the ecliptic. These parameters are also suggestive of a JFC-type impactor, although the less-common asteroidal type would be consistent with the observed geometry.


\subsection{Ensemble results}
In this section, we describe some of the properties of our observed ensemble of events and make some comparisons to the model populations described in Section \ref{sec:models}. 

\begin{figure}[b!]
\gridline{\fig{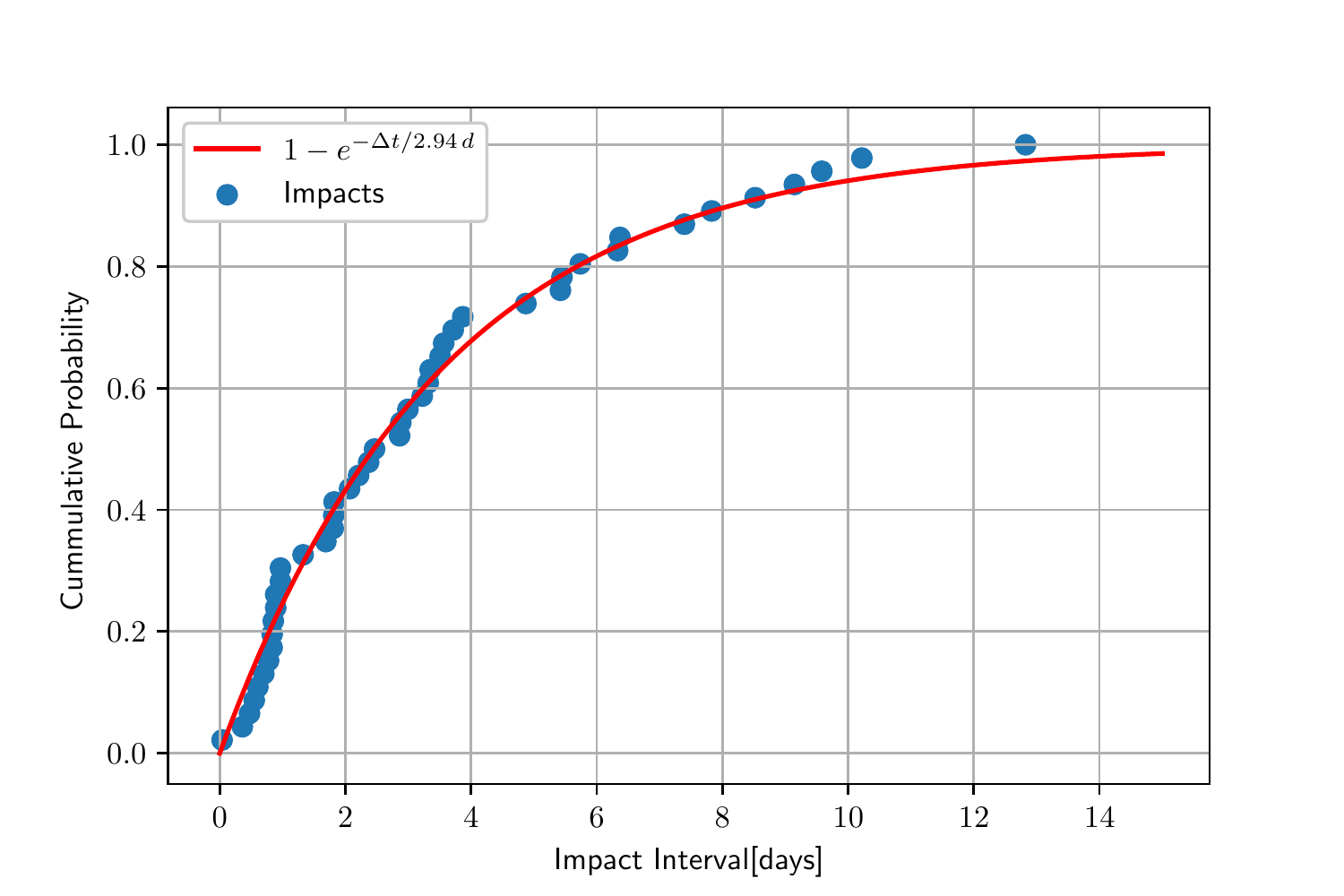}{0.5\textwidth}{}}
\vspace*{-8mm}
\caption{Cumulative distribution of observed time interval between events, taking into account gaps in the observations. The red curve is an exponential fit with a characteristic interval between events of $2.94\pm0.05\,$days. \label{fig:CDF_rate}}
\end{figure}

Micrometeoroid impact times are expected to be governed by a Poisson process characterized by a single rate parameter.  Figure \ref{fig:CDF_rate} shows the cumulative probability density of the observed time between events, which was computed from mission elapsed time by excising the times not included in our search. As is expected for a Poisson process, this distribution follows an exponential function, with a time between events of $2.94\pm0.05\,$days or a rate of $(124\pm2)\,\textrm{yr}^{-1}$.  This is consistent with the predictions in Section \ref{sec:models}.

From the observed impacts, we perform a hierarchical analysis to infer properties of the transferred momentum distribution and, assuming that the impact coefficient is similar for all impacts, the momentum distribution of the micrometeorite population.
We select only impacts with measured momenta $P>P_{\rm min}=1\ \mu N s$ as a threshold above which we assume 100\% detection efficiency and therefore neglect selection effects.
The marginalized posteriors of the momenta from the MCMC analysis described in section \ref{sec:MCMC} are approximated as Gaussian distributions, with mean and variance computed from the Markov chains.
The approximate posteriors become the data $d$ in a hierarchical analysis that compares three models for the probability density function of momenta: 
a single power law 
\begin{equation}
p(P;\alpha) = A( P / P_{\rm min} )^{-\alpha}
\end{equation}
a broken power law with fixed ``knee'' momentum $P^*$
\begin{equation}
p(P;\alpha,\beta) =
\begin{cases}
 A( P / P_{\rm min} )^{-\alpha} & \text{if $P \leq P^*$}\\
B( P / P_{\rm min} )^{-\beta} & \text{else }
\end{cases}
\end{equation}
and a three-parameter model $p(P;\alpha,\beta,P^*)$ with adjustable knee location, where $A$ and $B$ normalize the distributions.
A MCMC code is used to characterize each model, and from the maximum likelihood we compute the Bayesian Information Criterion (BIC,~\cite{schwarz1978}). The model that minimizes the BIC is preferred.

For these data, the one-parameter model is selected, with BIC scores of 48, 53, and 56  for $p(P;\alpha)$, $p(P;\alpha,\beta)$, and $p(P;\alpha,\beta,P^*)$, respectively. 
As a sanity check, we also confirmed that the marginalized posteriors $p(\alpha | d)$ and $p(\beta | d)$ were largely overlapping (or, the posterior $p(\alpha-\beta|d)$ peaks near zero) as would be expected in the case where the one-parameter model adequately described the data.
The spectral index is measured to be
$\alpha = 1.87^{2.08}_{1.68}$ 
quoted as the median, with upper and lower 90\% credible intervals from the posterior distribution function as super- and subscripts.

\begin{figure}[h!]
\gridline{\fig{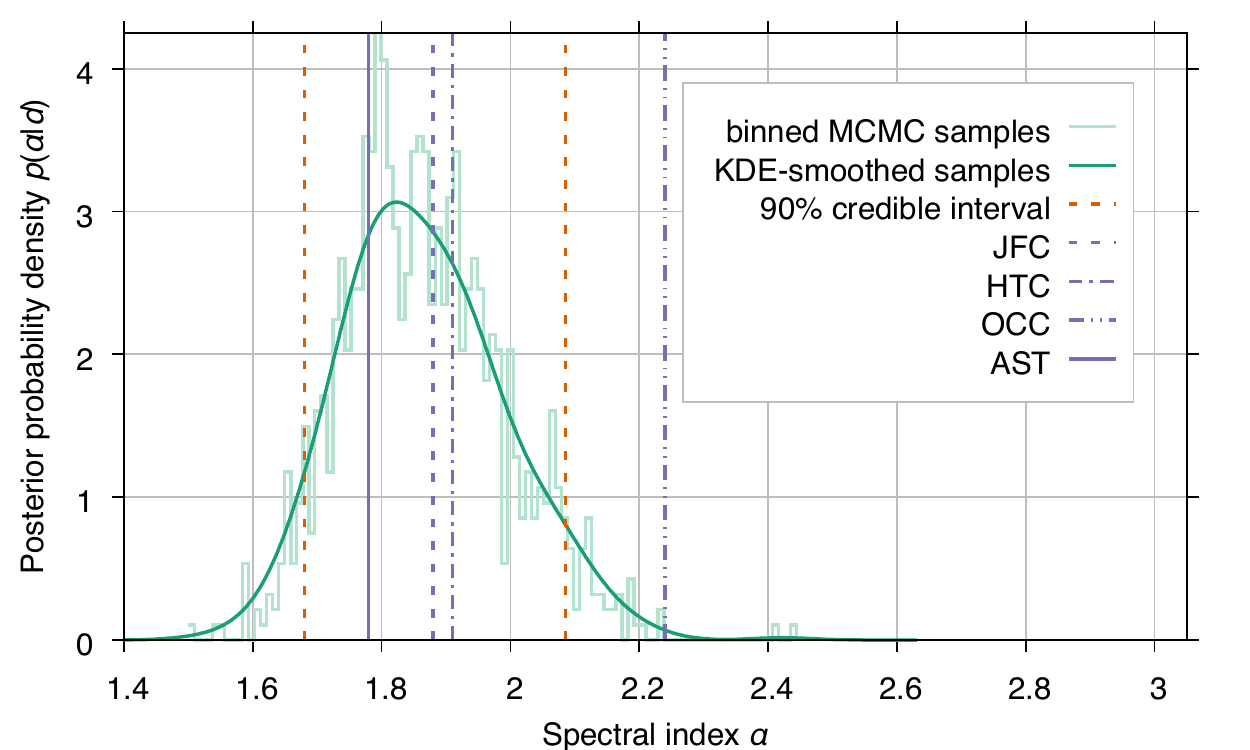}{0.5\textwidth}{}}
\vspace*{-8mm}
\caption{Inferred posterior distribution for momentum of underlying micrometeoroid population as a histogram of the chain samples (light blue-green) and a Kernel Density Estimate (dark blue-green) from a single-power-law model. The vertical dashed lines (orange) mark the 90\% credible intervals.  The four vertical lines (purple) are the best-fit power-law indices for the different micrometeorite progenitors as shown in Table~\ref{tab:popCDFfits}. The OCC model (dashed-double-dotted line) is disfavored by these observations. See text for details. \label{fig:PDF_P}}
\end{figure}

\begin{figure}[h!]
\gridline{\fig{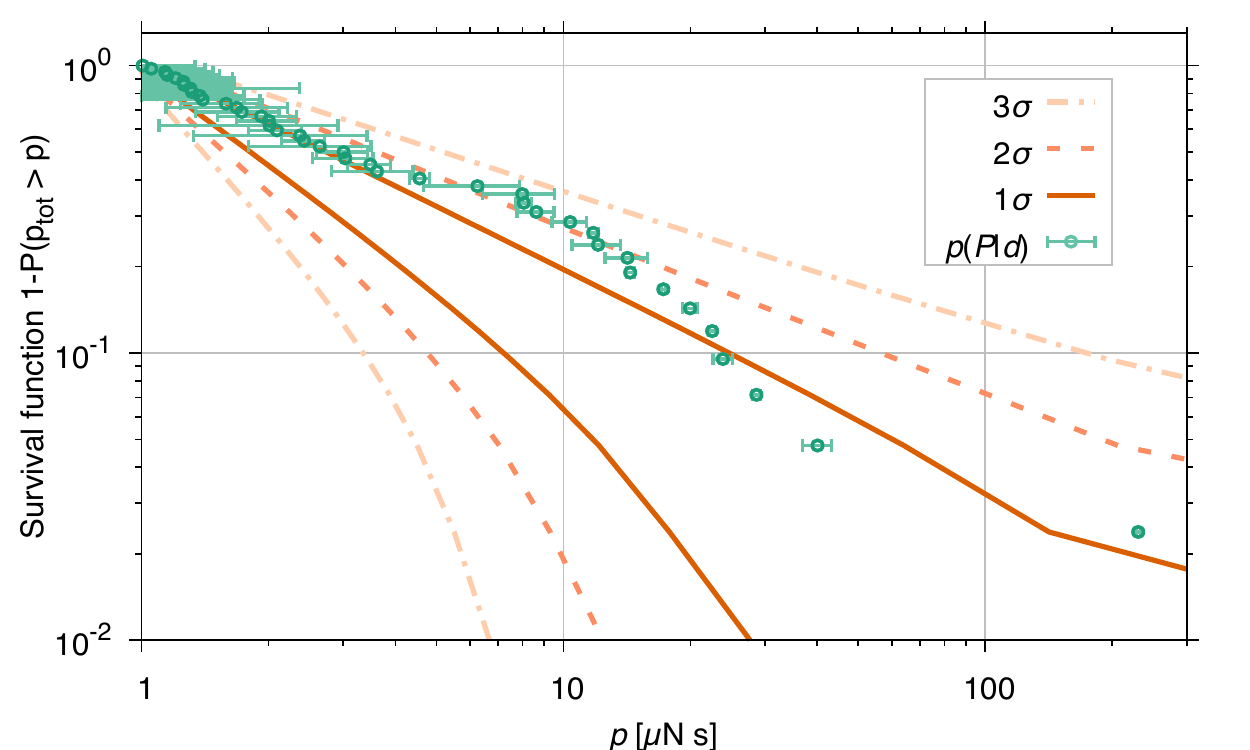}{0.5\textwidth}{}}
\vspace*{-8mm}
\caption{Cumulative distribution of impact momenta from the 54 measured events (green points with 90\% credible error bars from individual MCMC posteriors) as well as 1$\sigma$, 2$\sigma$, and 3$\sigma$ ranges for the underlying distribution generated using the kernel density estimate. These can be directly compared with the model distributions in Figure \ref{fig:popCDF}. See text for details. \label{fig:CDF_P}}
\end{figure}

Figure \ref{fig:PDF_P} shows the inferred posterior distribution as a histogram of the chain samples (light blue-green) and a kernel density estimate (dark blue-green) from the single-power-law model on $\alpha$. The vertical dashed lines (orange) mark the 90\% credible intervals.  The four vertical lines (purple) are the best-fit power-law indices for the different micrometeorite progenitors as shown in Table~\ref{tab:popCDFfits}. The OCC model (dashed-double-dotted line) is disfavored by these observations. Figure \ref{fig:CDF_P} shows a comparison of the cumulative distribution of impact momenta from the kernel density estimate (1$\sigma$, 2$\sigma$, and 3$\sigma$ intervals as solid, dashed, and dotted-dashed lines respectively) with the measured distribution of the individual impacts, shown in green with 90\% error bars from the individual MCMC posteriors. 

\begin{figure}[h!]
\gridline{\fig{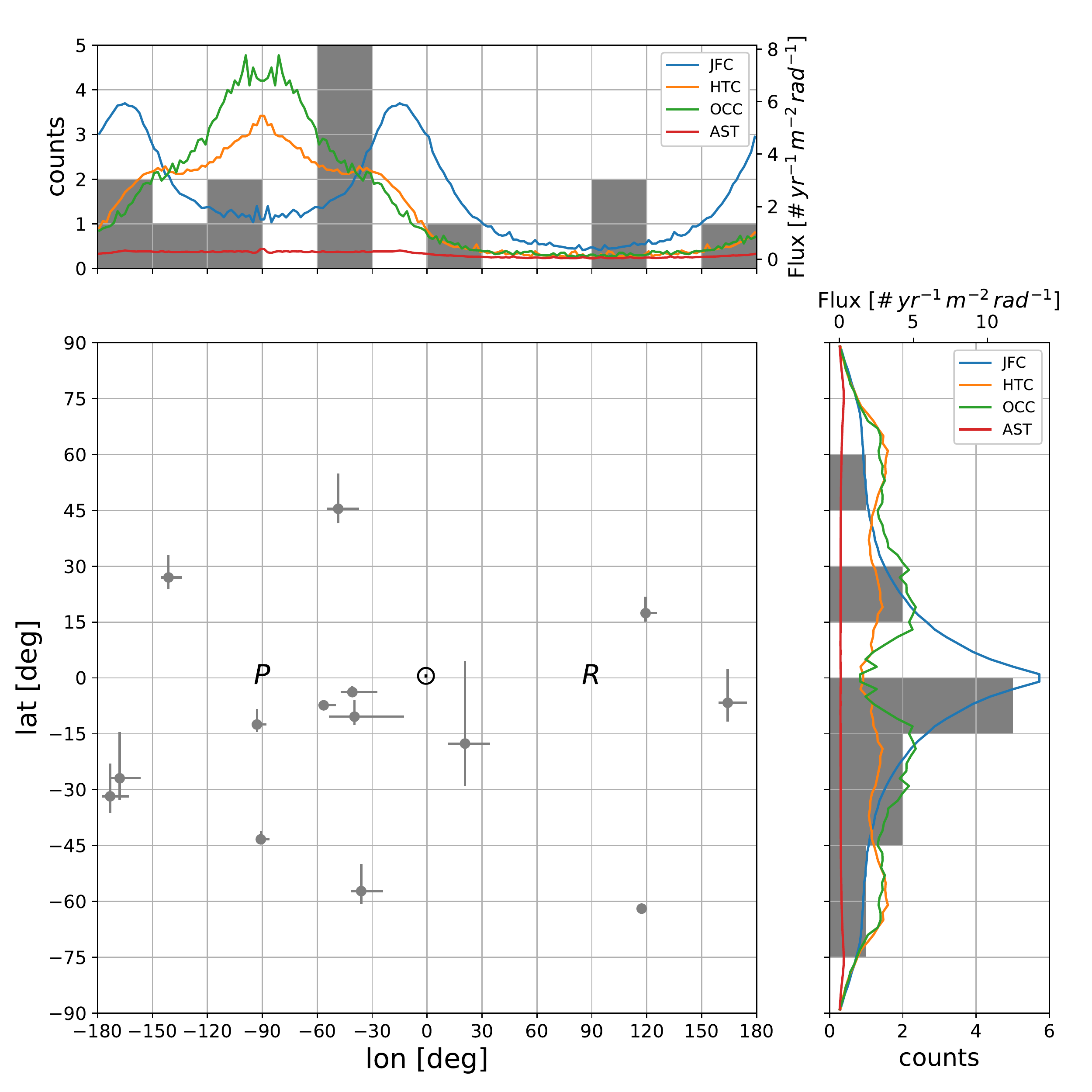}{0.5\textwidth}{}}
\vspace*{-8mm}
\caption{Comparison of sky distribution of localized events with population models. The main panel shows sky locations for the subset of (14) events that were localized to an area within 10\% of the sky, including error bars spanning a 68\% confidence level. The top panel shows in gray a histogram of the events in 15$^\circ$ bins of latitude, as well as the modeled flux distribution for impacts with momentum $\geq 1\,\mu$Ns for the JFC (blue), HTC (orange), OCC (green), and AST (red) populations. The right panel is similar to the top panel, but for latitude in 30$^\circ$ bins. See text for details. \label{fig:mapCompare}}
\end{figure}

A second way to distinguish between potential populations is to compare the distribution of events on the sky.  As mentioned in Section \ref{sec:sensitivity}, \emph{LPF}'s ability to localize events on the sky depends on detecting and measuring momentum transfer in multiple degrees of freedom. This is more likely to occur as the overall transferred momentum increases. Indeed, we find a correlation between total momentum and area of the 68\% confidence sky position of $\delta A\approx1.5\times10^4\,\textrm{deg}^2\,\left(P/1\,\mu\textrm{N}\right)^{-0.74}$. The main panel of figure \ref{fig:mapCompare} shows the measured sky position, assuming that the impact coefficient is independent of impact geometry, with 68\% error bars for the subset of 14 events for which the area of the 68\% confidence region on the sky is less than 4125$\,\textrm{deg}^2$ or 10\% of the sky. The top panel shows in gray a histogram of the events in 15$^\circ$ bins of latitude, as well as the modeled flux distribution for impacts with momentum $\geq 1\,\mu$Ns for the JFC (blue), HTC (orange), OCC (green), and AST (red) populations. The right panel is similar to the top panel, but for latitude in 30$^\circ$ bins.  While the limited number of well-localized events makes it difficult to quantitatively compare the data to the models, the distribution of events is suggestive of the JFC population, particularly in latitude. 

\section{Population Model Inference} \label{sec:model_inference}
To improve on the qualitative nature of the model comparison in Figure \ref{fig:mapCompare} and make a more qualitative statement about the agreement between the models in section \ref{sec:models} and \emph{LPF}'s observations, a hierachircal Bayesian model was developed that utilized the momentum and sky distribution of each population model to assess the likelihood that any particular step in the impact search chain was associated with an impact from a specific population.  This machinery was then applied to the \emph{entire} set of cleaned \emph{LPF} data, including segments for which no impact was positively identified (but excluding the few vetoed events). This is an important advantage, as nondetection of an event when a model predicts likely detections can be as important to model selection as detection of such events. The hierachircal model, which is described in detail in Appendix B, assumes that the underlying population of micrometeoroids is a mixture of a set of subpopulations and measures the posterior distributions of the relative contributions of these populations.   An important caveat in this analysis is that this analysis does not include an \emph{a priori} difference in the elastic coefficient $\beta$ between the two populations, which may be warranted owing to different compositions of the different populations.  While there is both a theoretical expectation and some experimental evidence that $\beta$ will decrease with increasing porosity~\citep{2018P&SS..164...91F}, there are not sufficient data at the impact velocities typical of the \emph{LPF} impacts to assign a specific ratio of $\beta$ between the different populations. Note that because we compare the relative likelihood of an impact arising from a particular source, our analysis will only be affected by errors in the \emph{difference} between impact ratios between populations rather than errors that are common to both populations.  We also expect that, based on the similarity of the momentum distributions in Figure~\ref{fig:popCDF} and the dissimilarity of the flux maps in Figure~\ref{fig:popMaps}, the population model selection will be driven primarily by impact direction rather than overall impact momenta. 

\begin{figure*}[ht!]
\gridline{\fig{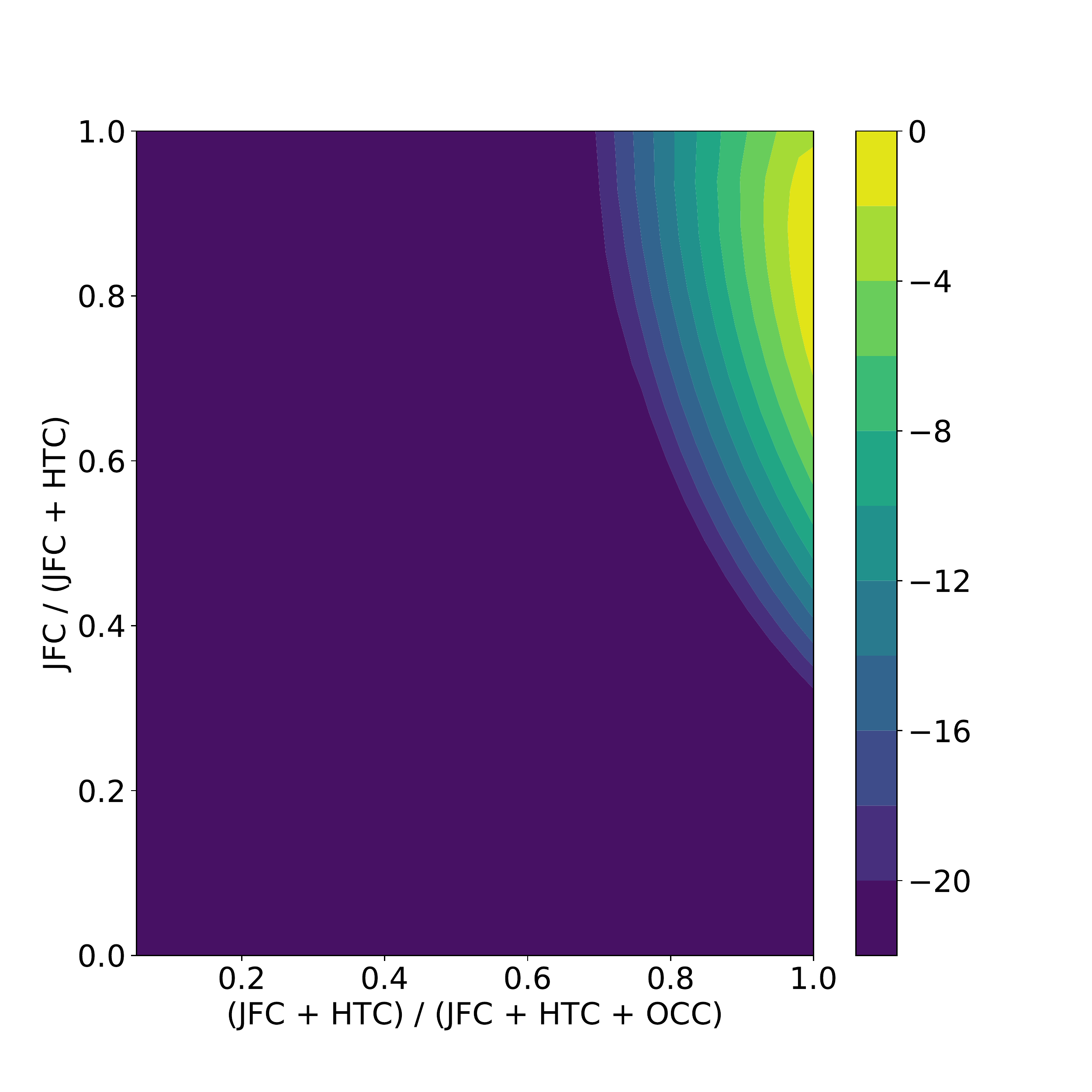}{\columnwidth}{(a) Sub-population ratio distribution with JFC, HTC and OCC components}
\fig{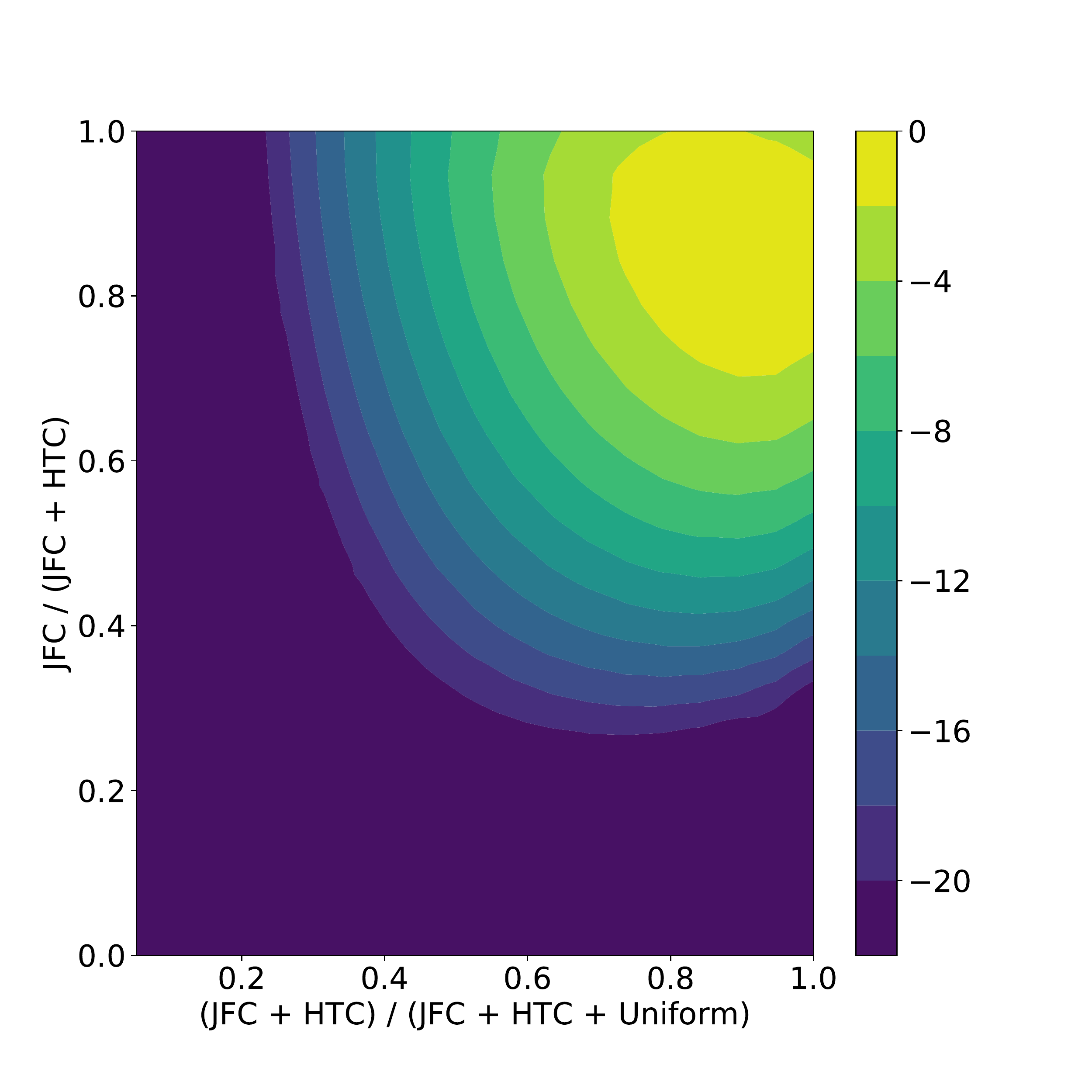}{\columnwidth}{(b) Sub-population ratio distribution with JFC, HTC and Uniform components}}
\caption{These figures show log Bayes factors for model comparisons with varying fractional contributions from our sub-population models. Differences of more than a few begin to be significant with a difference of 20 indicating very strong evidence.
  Panel a) shows most-probable relative rates of $\sim$80-90\% JFC with $\sim$10-20\% HTC micrometeoroids and no significant contribution from OCC.  Panel b) considers an alternative less-informed model leaving out the OCC sub-population, but allowing the possibility of an additional sub-population which is uniformly distributed across the sky. Models with JFC contributing to the majority the micrometeoroids remain strongly favored. \label{fig:bayes}}
\end{figure*}

In Figure~\ref{fig:bayes} we show the resulting Bayes factors for models composed of mixtures of the JFC, HTC, and OCC population models, as well as a Uniform-sky model which is used as a control.  The parameters of the hierarchical models are the fraction of net micrometeoroid flux assumed from each subpopulation, with the overall rate fixed to the observed rate. In the Figure~\ref{fig:bayes}(a), we consider models composed of a mixture of the JFC, HTC, and OCC subpopulations.  The result shows that models favoring predominantly JFC micrometeoroids are strongly favored while models with a large fraction of OCC micrometeoroids are especially disfavored.  The roughly 2:1 ratio of JFCs to HTCs predicted by the models outlined in section \ref{sec:models} lies in the region of maximum likelihood. The dominance of the combined JFC+HTC combinations over the OCC population is also consistent with these models so long as the threshold for observed impacts is greater than a few $\mu$N s.  Figure~\ref{fig:bayes}(b) shows results from a hierarchical model consisting of JFC, HTC, and Uniform-sky subpopulations. Models dominated by JFC micrometeoroids are again strongly favored.  

\section{Conclusions} \label{sec:conclusions}
We have presented a comprehensive analysis of micrometeoroid impacts detected by the \emph{LISA Pathfinder} spacecraft using a novel technique - direct measurement of the momentum transfer from an individual microscopic impactor to a spacecraft. This data set, although limited to a handful of events, provides an interesting new source of data for the zodiacal dust complex, an important component of our Solar System.  The population observed by \emph{LPF} is broadly consistent with standard models of the micrometeoroid population, suggesting that such models are appropriate for use in estimating hazards for spacecraft operating in the inner solar system.  A statistical comparison of our data set with model predictions favors models primarily composed of Jupiter-family Comets  with a potential smaller contribution from Halley-type Comets. This is broadly consistent with standard models of the zodiacal dust complex although our statistical evidence limited and effects of possible differences in the efficiency of momenta transfer for different impactors have not been assessed.  This same technique may be utilized by future precision-measurement missions, most notably the Laser Interferometer Space Antenna (LISA) itself, which based on this analysis, will observe many more micrometeoroid impacts due to its combination of more spacecraft, larger spacecraft, and longer observing time - providing an additional science benefit beyond the compelling science case for observing the universe in the milliHertz gravitational wave band.

\acknowledgments

This analysis was conducted under a 2017 NASA Science Innovation Fund awarded to Thorpe, Littenberg, Janches, and Baker. Slutsky acknowledges support of the NASA Astrophysics Division. Pagane and Hourihane acknowledge the support of the NASA Undergraduate Summer Internship program and Hourihane acknowledges the support of the National Science Foundation's Research Experience for Undergraduates program.

The data was produced by the LISA Pathfinder mission, which was part of the
space-science programme of the European Space Agency and also hosted the NASA Disturbance Reduction System payload, developed under the NASA New Millennium Program. 

The French contribution to LISA Pathfinder has been supported by the CNES (Accord Specific de projet
CNES 1316634/CNRS 103747), the CNRS, the Observatoire de Paris and the University
Paris-Diderot. E.~Plagnol and H.~Inchausp\'{e} would also like to acknowledge the
financial support of the UnivEarthS Labex program at Sorbonne Paris Cit\'{e}
(ANR-10-LABX-0023 and ANR-11-IDEX-0005-02).

The Albert-Einstein-Institut acknowledges the support of the German Space Agency,
DLR, in the development and operations of LISA Pathfinder. The work is supported by the Federal Ministry for Economic Affairs and Energy
based on a resolution of the German Bundestag (FKZ 50OQ0501 and FKZ 50OQ1601). 

The Italian contribution to LISA Pathfinder has been supported  by Agenzia Spaziale Italiana and Istituto
Nazionale di Fisica Nucleare.

The Spanish contribution to LISA Pathfinder has been supported by contracts AYA2010-15709 (MICINN),
ESP2013-47637-P, and ESP2015-67234-P (MINECO). M.~Nofrarias acknowledges support from
Fundacion General CSIC (Programa ComFuturo). F.~Rivas acknowledges an FPI contract
(MINECO).

The Swiss contribution to LISA Pathfinder was made possible by the support of the Swiss Space Office (SSO)
via the PRODEX Programme of ESA. L.~Ferraioli is supported by the Swiss National
Science Foundation.

The UK LISA Pathfinder groups wish to acknowledge support from the United Kingdom Space Agency
(UKSA), the University of Glasgow, the University of Birmingham, Imperial College,
and the Scottish Universities Physics Alliance (SUPA).

%


\software{LTPDA (\url{https://www.elisascience.org/ltpda/}), \:
MATPLOTLIB (\url{https://matplotlib.org/}),\:
HEALPY (\url{https://github.com/healpy/healpy}),\:
ligo.skymap (\url{https://lscsoft.docs.ligo.org/ligo.skymap})}



 \appendix
 \section{List of Impact Events in \emph{LISA Pathfinder}}
The catalog of \emph{LPF} impact events is reported in the table below. Column (1) is the GPS time stamp of the event. Column (2) is the median transferred momentum ($\rho_{med}$) with 95\% confidence intervals. Column 3 is the median intrinsic particle momentum ($\Gamma$) assuming an impact coefficient ($\beta=3$), with an error including the statistical error in $\rho$ as well as an uncertainty in the impact coefficient $(1\leq \beta \leq 5)$. For impacts with a greater than 75\% probability of impacting on a particular face of the spacecraft, the spacecraft face is identified in Column (4). For impacts with an inferred 95\% confidence sky location of less than 4100 $deg^2$ (10\% of the sky), the 95\% error area as well as the impact direction in both spacecraft and Sun-synchronous ecliptic coordinates is reported in Columns (5)-(9). The location of \emph{LPF} in its orbit at the time of the impact is provided in EME2000 (J2000) coordinates in Columns (10)-(12).

			\begingroup
			\renewcommand\arraystretch{2}
			\begin{longtable}{|c|c|c|c|c|c|c|c|c|c|c|c|}
				\multicolumn{9}{c}
				{{\bfseries \tablename\  \thetable{}}}\\
				\hline \multicolumn{1}{|c|}{\textbf{GPS}}  & 
				\multicolumn{1}{|c|}{\bf{$\rho_{med}$ [$\mu Ns$]}} & 
				\multicolumn{1}{|c}{\bf{$\Gamma$ [$\mu Ns$]}} & 
				\multicolumn{1}{|c|}{\textbf{Face}} &
				\multicolumn{1}{|c|}{\textbf{Sky Area}} &
				\multicolumn{1}{|c|}{\textbf{$Lat_{SC}$}} &
				\multicolumn{1}{|c|}{\textbf{$Lon_{SC}$}} &
				\multicolumn{1}{|c|}{\textbf{$Lat_{SSE}$}} &
				\multicolumn{1}{|c|}{\textbf{$Lon_{SSE}$}} &
				\multicolumn{1}{|c|}{\textbf{$LPF_X$}} &
				\multicolumn{1}{|c|}{\textbf{$LPF_Y$}} &
				\multicolumn{1}{|c|}{\textbf{$LPF_Z$}} \\
				\hline
			\endfirsthead
			
			\multicolumn{9}{c}
				{{\bfseries \tablename\  \thetable{} -- continued from previous page}} \\
			\hline \multicolumn{1}{|c|}{\textbf{GPS}}  & 
				\multicolumn{1}{|c|}{\bf{$\rho_{med}$ [$\mu Ns$]}} & 
				\multicolumn{1}{|c}{\bf{$\Gamma$ [$\mu Ns$]}} & 
				\multicolumn{1}{|c|}{\textbf{Face}} &
				\multicolumn{1}{|c|}{\textbf{Sky Area}} &
				\multicolumn{1}{|c|}{\textbf{$Lat_{SC}$}} &
				\multicolumn{1}{|c|}{\textbf{$Lon_{SC}$}} &
				\multicolumn{1}{|c|}{\textbf{$Lat_{SSE}$}} &
				\multicolumn{1}{|c|}{\textbf{$Lon_{SSE}$}} &
				\multicolumn{1}{|c|}{\textbf{$LPF_X$}} &
				\multicolumn{1}{|c|}{\textbf{$LPF_Y$}} &
				\multicolumn{1}{|c|}{\textbf{$LPF_Z$}} \\
				\hline
			\endhead
			
			\hline \multicolumn{9}{|r|}{{Continued on next page}} \\ \hline
			\endfoot

			\hline
			\endlastfoot
	1144229908 & $17.2^{+0.4}_{-0.3}$ & $ 5.7^{+11.9}_{-2.4}$ & +y+y & 1729 & -7 & -7 & -57 & -39 & 1.09 & 0.55 & -0.05 \\
	1146429822 & $ 1.7^{+3.1}_{-0.6}$ & $ 0.6^{+4.2}_{-0.3}$ & - & - & - & - & - & - & 0.45 & 1.24 & 0.56 \\
	1147442122 & $ 0.7^{+0.5}_{-0.5}$ & $ 0.2^{+1.0}_{-0.2}$ & - & - & - & - & - & - & 0.14 & 1.34 & 0.77 \\
	1147453726 & $14.4^{+0.8}_{-0.4}$ & $ 4.8^{+10.4}_{-2.0}$ & +x+x & 3438 & -2 & 162 & 45 & -56 & 0.14 & 1.34 & 0.77 \\
	1147693044 & $ 0.9^{+0.9}_{-0.3}$ & $ 0.3^{+1.5}_{-0.2}$ & - & - & - & - & - & - & 0.08 & 1.35 & 0.81 \\
	1147741578 & $ 2.0^{+0.8}_{-0.3}$ & $ 0.7^{+2.2}_{-0.3}$ & - & - & - & - & - & - & 0.06 & 1.35 & 0.82 \\
	1149475988 & $ 1.0^{+1.1}_{-0.3}$ & $ 0.3^{+1.8}_{-0.2}$ & - & - & - & - & - & - & -0.26 & 1.36 & 1.00 \\
	1150511110 & $ 3.5^{+1.7}_{-1.2}$ & $ 1.2^{+4.1}_{-0.7}$ & - & - & - & - & - & - & -0.35 & 1.33 & 1.03 \\
	1151901050 & $ 0.2^{+0.5}_{-0.1}$ & $ 0.1^{+0.7}_{-0.0}$ & - & - & - & - & - & - & -0.42 & 1.35 & 1.00 \\
	1153404058 & $ 2.9^{+1.3}_{-0.3}$ & $ 1.0^{+3.2}_{-0.4}$ & - & - & - & - & - & - & -0.48 & 1.37 & 0.87 \\
	1153750663 & $19.9^{+1.7}_{-1.3}$ & $ 6.6^{+15.0}_{-2.9}$ & +z+z & 2585 & 18 & 158 & -31 & -172 & -0.51 & 1.38 & 0.83 \\
	1154024345 & $ 8.6^{+1.8}_{-1.6}$ & $ 2.9^{+7.6}_{-1.5}$ & +x+y & 3857 & -7 & 128 & -7 & 156 & -0.54 & 1.38 & 0.79 \\
	1154963503 & $ 2.4^{+0.8}_{-0.3}$ & $ 0.8^{+2.4}_{-0.4}$ & - & - & - & - & - & - & -0.66 & 1.35 & 0.64 \\
	1155461605 & $ 0.5^{+1.3}_{-0.3}$ & $ 0.2^{+1.6}_{-0.1}$ & - & - & - & - & - & - & -0.73 & 1.31 & 0.54 \\
	1155558407 & $ 1.6^{+1.3}_{-0.8}$ & $ 0.5^{+2.4}_{-0.4}$ & - & - & - & - & - & - & -0.74 & 1.30 & 0.52 \\
	1155637974 & $12.1^{+3.0}_{-3.2}$ & $ 4.0^{+11.1}_{-2.3}$ & +z+z & 1786 & 68 & -87 & -4 & -39 & -0.76 & 1.29 & 0.50 \\
	1155677822 & $ 2.4^{+1.7}_{-2.3}$ & $ 0.8^{+3.3}_{-0.8}$ & +z+z & - & - & - & - & - & -0.76 & 1.29 & 0.50 \\
	1155891413 & $ 0.7^{+0.3}_{-0.2}$ & $ 0.2^{+0.8}_{-0.1}$ & - & - & - & - & - & - & -0.80 & 1.26 & 0.45 \\
	1155985559 & $23.8^{+2.6}_{-2.1}$ & $ 7.9^{+18.5}_{-3.6}$ & +z+z & 84 & 87 & -112 & -7 & -58 & -0.82 & 1.25 & 0.43 \\
	1156020427 & $ 0.9^{+3.0}_{-0.7}$ & $ 0.3^{+3.5}_{-0.2}$ & +z+z & - & - & - & - & - & -0.83 & 1.25 & 0.42 \\
	1156063801 & $ 1.0^{+0.7}_{-0.8}$ & $ 0.3^{+1.4}_{-0.3}$ & - & - & - & - & - & - & -0.83 & 1.24 & 0.41 \\
	1156115516 & $ 3.0^{+1.0}_{-0.9}$ & $ 1.0^{+3.0}_{-0.6}$ & +z+z & 1873 & 77 & -105 & -11 & -53 & -0.84 & 1.23 & 0.40 \\
	1156188047 & $ 0.5^{+1.2}_{-0.4}$ & $ 0.2^{+1.5}_{-0.1}$ & - & - & - & - & - & - & -0.86 & 1.22 & 0.39 \\
	1156255314 & $ 0.6^{+2.8}_{-0.3}$ & $ 0.2^{+3.2}_{-0.1}$ & - & - & - & - & - & - & -0.87 & 1.21 & 0.37 \\
	1157966718 & $ 1.1^{+1.3}_{-0.4}$ & $ 0.4^{+2.0}_{-0.2}$ & - & - & - & - & - & - & -1.14 & 0.70 & -0.04 \\
	1159736213 & $ 0.9^{+1.5}_{-0.6}$ & $ 0.3^{+2.0}_{-0.2}$ & +z+z & - & - & - & - & - & -1.15 & -0.18 & -0.41 \\
	1159808666 & $230.3^{+4.8}_{-5.8}$ & $76.8^{+158.3}_{-31.9}$ & +x+y & 430 & 4 & 101 & -62 & 116 & -1.14 & -0.21 & -0.42 \\
	1159869088 & $ 6.4^{+2.8}_{-3.4}$ & $ 2.1^{+7.1}_{-1.5}$ & +z+z & 2645 & 66 & 3 & -18 & 6 & -1.13 & -0.25 & -0.43 \\
	1164719570 & $ 0.6^{+0.6}_{-0.3}$ & $ 0.2^{+1.0}_{-0.1}$ & - & - & - & - & - & - & 0.06 & -1.62 & -0.29 \\
	1166268578 & $ 8.0^{+3.1}_{-2.8}$ & $ 2.7^{+8.4}_{-1.6}$ & - & - & - & - & - & - & 0.20 & -1.65 & -0.23 \\
	1166337501 & $ 1.6^{+1.1}_{-0.4}$ & $ 0.5^{+2.2}_{-0.3}$ & - & - & - & - & - & - & 0.21 & -1.65 & -0.23 \\
	1166805122 & $ 0.5^{+1.1}_{-0.4}$ & $ 0.2^{+1.4}_{-0.1}$ & - & - & - & - & - & - & 0.23 & -1.65 & -0.24 \\
	1166921605 & $28.6^{+1.2}_{-0.9}$ & $ 9.5^{+20.3}_{-4.0}$ & +y-x & 1716 & 19 & 13 & -12 & -91 & 0.24 & -1.65 & -0.24 \\
	1166995369 & $ 0.8^{+0.9}_{-0.3}$ & $ 0.3^{+1.4}_{-0.2}$ & - & - & - & - & - & - & 0.25 & -1.64 & -0.25 \\
	1167307196 & $22.5^{+0.8}_{-0.7}$ & $ 7.5^{+15.8}_{-3.1}$ & +x+x & 2149 & -7 & 150 & 17 & 114 & 0.26 & -1.64 & -0.26 \\
	1167613479 & $ 0.9^{+1.0}_{-0.3}$ & $ 0.3^{+1.6}_{-0.2}$ & - & - & - & - & - & - & 0.28 & -1.62 & -0.28 \\
	1167654180 & $10.3^{+2.1}_{-1.5}$ & $ 3.4^{+8.9}_{-1.7}$ & - & - & - & - & - & - & 0.28 & -1.62 & -0.28 \\
	1167944728 & $ 4.5^{+0.6}_{-0.3}$ & $ 1.5^{+3.7}_{-0.7}$ & -y-y & - & - & - & - & - & 0.30 & -1.61 & -0.30 \\
	1168061759 & $ 3.5^{+0.9}_{-0.7}$ & $ 1.2^{+3.2}_{-0.6}$ & -y-y & - & - & - & - & - & 0.30 & -1.60 & -0.31 \\
	1168267680 & $ 1.2^{+1.0}_{-0.3}$ & $ 0.4^{+1.8}_{-0.2}$ & - & - & - & - & - & - & 0.31 & -1.59 & -0.33 \\
	1170979672 & $ 1.8^{+1.2}_{-0.4}$ & $ 0.6^{+2.4}_{-0.3}$ & - & - & - & - & - & - & 0.68 & -1.22 & -0.60 \\
	1171012017 & $ 2.5^{+2.2}_{-1.1}$ & $ 0.8^{+3.9}_{-0.5}$ & - & - & - & - & - & - & 0.68 & -1.22 & -0.60 \\
	1173291241 & $ 1.9^{+0.9}_{-0.3}$ & $ 0.6^{+2.2}_{-0.3}$ & - & - & - & - & - & - & 1.12 & -0.38 & -0.55 \\
	1176914535 & $ 1.0^{+1.0}_{-0.3}$ & $ 0.3^{+1.7}_{-0.2}$ & - & - & - & - & - & - & 0.76 & 1.10 & 0.43 \\
	1176917343 & $ 1.1^{+1.2}_{-0.3}$ & $ 0.4^{+1.9}_{-0.2}$ & - & - & - & - & - & - & 0.76 & 1.10 & 0.43 \\
	1177956916 & $ 1.3^{+1.1}_{-0.3}$ & $ 0.4^{+1.9}_{-0.2}$ & - & - & - & - & - & - & 0.48 & 1.27 & 0.70 \\
	1178035038 & $40.2^{+5.8}_{-6.6}$ & $13.4^{+32.6}_{-6.7}$ & -y+x & 168 & -83 & -63 & -43 & -91 & 0.46 & 1.27 & 0.72 \\
	1178120384 & $ 1.5^{+1.0}_{-0.3}$ & $ 0.5^{+2.0}_{-0.3}$ & - & - & - & - & - & - & 0.44 & 1.28 & 0.73 \\
	1178197245 & $ 1.2^{+1.1}_{-0.3}$ & $ 0.4^{+1.9}_{-0.2}$ & - & - & - & - & - & - & 0.43 & 1.29 & 0.75 \\
	1178251226 & $11.7^{+0.9}_{-0.3}$ & $ 3.9^{+8.7}_{-1.6}$ & -y-y & - & - & - & - & - & 0.41 & 1.29 & 0.76 \\
	1179167273 & $14.0^{+3.9}_{-2.5}$ & $ 4.7^{+13.2}_{-2.4}$ & +x+y & 3015 & 8 & 84 & 27 & -142 & 0.23 & 1.33 & 0.93 \\
	1179493289 & $ 8.0^{+0.8}_{-0.5}$ & $ 2.7^{+6.2}_{-1.2}$ & +y-x & 3864 & -1 & 25 & -27 & -173 & 0.18 & 1.34 & 0.97 \\
	1180613326 & $ 1.2^{+1.5}_{-0.4}$ & $ 0.4^{+2.3}_{-0.2}$ & - & - & - & - & - & - & 0.02 & 1.33 & 1.08 \\
	1181272382 & $ 1.0^{+1.0}_{-0.3}$ & $ 0.3^{+1.7}_{-0.2}$ & - & - & - & - & - & - & -0.02 & 1.33 & 1.11 \\
	\hline
\end{longtable} 
\endgroup

\pagebreak
\section{Description of Population Model selection tool}
Using hierarchical Bayesian analysis, we can piggyback on our Bayesian treatment of impacts to make inferences about the populations producing those impacts. The hierarchical analysis begins by considering a broader model including both the population and impact processes, which we may jointly parameterize by $\theta=\{\theta_P,\theta_I\}$, combining ``population'' parameters and ``impact'' parameters, respectively.  For the joint model, Bayes's theorem looks like
\begin{equation}
  p(\theta_I,\theta_P|D)=\frac{p(D|\theta_I,\theta_P)p(\theta_I|\theta_P)p(\theta_P)}{p(D)}.
\end{equation}
where $D=\{D_\alpha\}$ is the combined full set of \emph{LPF} data segments considered here and $\theta_I$ is abstractly encompasses impact parameters across the full data set.

Here we are primarily interested in $\theta_P$, describing the population models, as in Sec.~\ref{sec:models}, so we marginalize over $\theta_I$.  This provides a Bayesian framework for population inference
\begin{eqnarray}
  p(\theta_P|D)&=&\frac{ p(D|\theta_P)p(\theta_P)}{p(D)}\label{eq:HierBayesThm}\\
  p(D|\theta_P)&=&\int p(D|\theta_I)p(\theta_I|\theta_P)d\theta_I\label{eq:metaLike}.
\end{eqnarray}
The second line expresses the effective likelihood function that we need for the population model inference analysis.

In practice, we assume that impacts for each data segment are independent, so that
\begin{eqnarray}
  \ln {p(D|\theta_P)}&=&\sum_\alpha \ln\int {p(D_\alpha|\theta_{I\alpha})}p(\theta_{I\alpha}|\theta_P)d\theta_{I\alpha}\\
  &=&\sum_\alpha \ln\hat{\mathrm{E}}\left[\frac{p(\theta_{I\alpha}|\theta_P)}{\hat p(\theta_I)}\right]+\mathrm{const}\label{eq:metaLikeSegmented}.
\end{eqnarray}
The contribution from each data segment is expressed as the expected value (with respect to the population-model-independent impact posterior distribution) of the ratio of the model-informed impact prior $p(\theta_{I\alpha}|\theta_P)$ to the uninformed prior $\hat p(\theta_{I})$ that was assumed in our impact analysis.
This assumes that the model-informed prior has no support outside the region of support for $\hat p(\theta_{I})$.
The crucial step to complete the computation of Equation (\eqref{eq:HierBayesThm}) is to estimate these expected values using the usual Bayesian approach of averaging over a posterior distributed sample, which we have already constructed via MCMC.

Putting all this together for our trans-dimensional MCMC model allowing zero or one impacts per segment, we get
\begin{eqnarray}
  \ln p(\theta_P|D)&=& \sum_\alpha\ln\Big[ (1-\epsilon_\alpha)(1-r_\alpha(\theta_P))+r_\alpha(\theta_P)\frac{\epsilon_\alpha}{N_{\mathrm{det}\,\alpha}}\sum_{s\in S_{\alpha,k}}\hat r_\alpha(\psi_s|\theta_P)\Big]+\ln p(\theta_P)+\mathrm{const}\label{eq:HierPost}
\end{eqnarray}
where $\epsilon_\alpha$ is the MCMC impact probability, and $S_{\alpha}$ is the set of $N_{\mathrm{det}\,\alpha}$ MCMC samples with impacts for segment $\alpha$, $r_\alpha(\theta_P)$ is the probability of an impact during this data segment for population model parameters $\theta_P$ and $\hat r_\alpha(\psi|\theta_P)$ is the informed prior probability of impact parameters $\psi$ assuming an impact.

 We write the time-segment \emph{LPF}-frame rate $r_\alpha(\psi,\theta_P)=\hat r_\alpha(\psi|\theta_P)r_\alpha(\theta_P)$ in terms of the physical micrometeoriod fluxes $F(\bar\theta,\bar\phi,\bar P,\theta_P)$ by
\begin{align}
  r_\alpha(\psi,\theta_P)&=T_\alpha A_{LPF}\frac{\partial(\bar\theta,\bar\phi,\bar P)}{\partial\psi}F(\bar\theta(\psi,t_\alpha),\bar\phi(\psi,t_\alpha),\bar P(\psi),\theta_P)
\end{align}
where $T_\alpha$ is the duration of the observation segment in time, $A_{LPF}$ is the spacecraft area, and the derivative factor is the Jacobian of the transformation from \emph{LPF} parameters to the population model dimensions $\{\bar\theta,\bar\phi,\bar P\}$ at observation time $t_\alpha$.

We assume that an overall population model consisting of some linear combination of the JFC, HTC, and OCC fluxes introduced in Sec.~\ref{sec:models} together with a naive baseline model assuming directionally uniform flux inversely proportional to impact momentum. Having constrained the overall rate, we replace $r_\alpha(\theta_P)$ in Equation (\eqref{eq:HierPost}) with our \emph{a posteriori} per-segment rate estimate. 
We normalize the flux from each of these subpopulations to the fixed overall rate, and then we combine these linearly, writing
\begin{align}
  F(\bar\theta,\bar\phi,\bar P,\theta_P)&=\sum_{\lambda} c_\lambda \hat F_\lambda(\bar\theta,\bar\phi,\bar P).\nonumber
\end{align}
With the overall rate fixed, the remaining population model parameters fractional subpopulation weights $\theta_P\equiv{\hat c_\lambda}$ normalized by $\sum\hat c_\lambda=1$. In practice, in Fig.~\ref{fig:bayes} we consider two versions of such a master model, each time incorporating the JFC and HTC subpopulations, but alternatively considering the OCC or the Uniform-sky populations as a third component.




\bibliography{Bibliography2}


\listofchanges

\end{document}